\documentclass[aps,prc,twocolumn,10pt, superscriptaddress, showpacs, floatfix]{revtex4-1}

\usepackage{amssymb,epsfig}
\usepackage{bbold}

\newcommand{\be}{\begin{equation}}
\newcommand{\ee}{\end{equation}}
\newcommand{\bea}{\begin{eqnarray}}
\newcommand{\eea}{\end{eqnarray}}

\begin{document}

\title{Many-body theory for quasiparticle states in superfluid fermionic systems}
\author{Elena Litvinova}
\affiliation{Department of Physics, Western Michigan University, Kalamazoo, MI 49008, USA}
\affiliation{National Superconducting Cyclotron Laboratory, Michigan State University, East Lansing, MI 48824, USA}
\author{Yinu Zhang}
\affiliation{Department of Physics, Western Michigan University, Kalamazoo, MI 49008, USA}


\date{\today}

\begin{abstract}
We present a formalism for the fermionic quasiparticle propagator in a superfluid fermionic system. Starting from a general many-body hamiltonian confined by the two-body {\it instantaneous} interaction, the equation of motion for the fermionic propagator is obtained in the Dyson form. Before making any approximation, the interaction kernel is found to be decomposed into the static and dynamical (time-dependent) contributions, while the latter translates to the energy-dependent and the former maps to the energy-independent terms in the energy domain. The three-fermion correlation function being the heart of the dynamical part of the kernel is factorized into the two-fermion and one-fermion ones. With the relaxed particle number constraint, the normal propagator is coupled to the anomalous one via both the static and dynamical kernels, that is formalized by introducing the generalized quasiparticle propagator of the Gor'kov type.  The dynamical kernel in the factorized form is associated with the quasiparticle-vibration coupling (QVC) with the vibrations unifying both the normal and pairing phonons.  The QVC vertices are related to the variations of the Hamiltonian of the Bogoliubov quasiparticles, which can be obtained by the finite amplitude method. 
\end{abstract}
\maketitle

\section{Introduction} 
The nuclear many-body problem attracts tremendous attention and effort for decades. Yet, the quantitative description of atomic nuclei calls for further developments, both in the theoretical and computational aspects. The predictive power and accuracy required for modern applications, such as nuclear astrophysics, neutrino physics, and searches beyond the standard model in the nuclear sector, are still much higher than can be offered even by the best advanced nuclear models and computation. 

One of the most powerful methods to study quantum many-body systems is the Green function method, because various Green functions, or propagators, being particular cases of a larger class of correlation functions, are directly related to well-defined observables. In the physics of atomic nuclei, the single-nucleon propagators are linked to the energies of odd-particle systems and spectroscopic factors, which can be extracted from, e.g., transfer or knock-out reactions, and two-nucleon propagators are associated with nuclear response to external probes of electromagnetic, strong or weak character. 

In the many-body theory, these types of propagators are connected through the equations of motion (EOM). For instance, the EOM for one-fermion propagator is coupled to that for higher-rank propagators via the dynamical kernel \cite{Rowe1968,Schuck1976,AdachiSchuck1989,Danielewicz1994,DukelskyRoepkeSchuck1998}. The importance of the latter couplings in nuclear systems was realized also in phenomenological approaches of the nuclear field theory (NFT) \cite{BohrMottelson1969,BohrMottelson1975,Broglia1976,BortignonBrogliaBesEtAl1977,BertschBortignonBroglia1983,LitvinovaRing2006} and the quasiparticle-phonon models (QPM) \cite{Soloviev1992,Malov1976} without, however, the explicit link to the underlying bare nucleon-nucleon interaction. The EOM method for approximating fermionic propagators is actively employed in quantum chemistry and condensed matter physics
\cite{Tiago2008,Martinez2010,Sangalli2011,SchuckTohyama2016,Olevano2018}, see 
the recent review \cite{Schuck2021rev} devoted to its systematic assessment.

In this work we aim at an extension of the single-fermion EOM to the superfluid case. Although some versions of such an extension are available in the literature, they are either based on the phenomenological assumptions about the dynamical kernel \cite{VanderSluys1993,Avdeenkov1999,Avdeenkov1999a,BarrancoBrogliaGoriEtAl1999,BarrancoBortignonBrogliaEtAl2005,Tselyaev2007,LitvinovaAfanasjev2011,Litvinova2012a,AfanasjevLitvinova2015,IdiniPotelBarrancoEtAl2015} or use perturbation theory to approximate it \cite{Soma2011,Soma2013,Soma2014a,Soma2021}. 
Another class of advanced models based on the equation of
motion phonon method (EMPM) treats the nuclear many-body
problem in terms of coupling between the Tamm-Dancoff phonons and quasiparticles. Furthermore, this method provides a
recipe for an "ab initio" determination of the
particle(hole)-vibration coupling \cite{DeGregorio2016} and for
coupling to complex configurations, such as two and three phonons.  It extends to the 
particle-hole vibrations \cite{Bianco2012} as well as
the quasiparticle (normal and pairing) vibrational modes \cite{DeGregorio2016a}.

Here we elaborate on (i) a continuous derivation of the EOM for the fermionic quasiparticle propagator starting from the underlying many-body fermionic Hamiltonian with the bare two-fermion interaction and (ii) on the non-perturbative treatment of the dynamical kernel of the resulting EOM on equal footing with the superfluid pairing correlations. We walk the reader throughout the detailed formalism to provide a clear understanding of the origin of the many-body effects, in particular, of the emergent collective phenomena in strongly-correlated superfluid media. The model-independent nature of the exact EOM's should allow practitioners to relate different many-body approaches to each other and to evaluate the accuracy of model approximations. In this article, we purposefully do not present numerical calculations to fully focus on the formal aspects of the theory as it is done, for instance, in Refs. \cite{Tselyaev2007,Soma2011}.

The article is organized as follows. In Section \ref{Propagators} we introduce notations and definitions of the fermionic propagators in correlated media.  Section \ref{EOM1} is devoted to the equations of motion and associated observables: Subsection \ref{normal} reviews briefly the introductory piece of the EOM formalism regarding the normal (non-superfluid) single-fermion propagator in a strongly-coupled regime, which was set up in detail in Ref. \cite{LitvinovaSchuck2019}; Subsection  \ref{superfluid} extends the framework to the superfluid case, i.e., to additional anomalous fermionic propagators; Subsection \ref{SE_qp} elaborates on the dynamical kernels of the obtained EOM's and on the unified treatment of the normal and pairing phonons in these kernels, which happen to be components of one unified dynamical kernel in the quasiparticle world;  Subsection \ref{SFAmpl} relates the observed strength functions to the quasiparticle transition amplitudes. Section \ref{PVCvert} establishes a link between the quasiparticle-vibration coupling (QVC) vertices in the dynamical kernel and the variations of the quasiparticle Hamiltonian in the context of the finite amplitude method (FAM) \cite{NakatsukasaInakuraYabana2007,Kortelainen2015,Bjelcic2020}. Section \ref{summary} contains a summary of the presented work, and the Appendix relates the formalism of this work to the Gor'kov theory, which turns out to be the static limit of the proposed approach. 


\section{Fermionic propagators in a correlated medium}
\label{Propagators}

We aim at developing a framework, where the many-body Hamiltonian serves as the only input, which determines uniquely all the properties of the system of interacting fermions. 
Thus, the starting point is the fermionic Hamiltonian, here in a non-relativistic form:
\be
H = H^{(1)} + V^{(2)} + W^{(3)} + ...,
\label{Hamiltonian}
\ee
where the operator $H^{(1)}$ is the one-body contribution:
\be
H^{(1)} = \sum_{12} t_{12} \psi^{\dag}_1\psi_2 + \sum_{12}v^{(MF)}_{12}\psi^{\dag}_1\psi_2 \equiv \sum_{12}h_{12}\psi^{\dag}_1\psi_2
\label{Hamiltonian1}
\ee
with matrix elements $h_{12}$ which, in general, combine the kinetic energy $t$ and the mean-field $v^{(MF)}$ part of the interaction. The  two-body sector associated with the two-fermion interaction is described by the operator $V^{(2)}$:
\be
V^{(2)} = \frac{1}{4}\sum\limits_{1234}{\bar v}_{1234}{\psi^{\dagger}}_1{\psi^{\dagger}}_2\psi_4\psi_3,
\label{Hamiltonian2}
\ee
while $W^{(3)}$ represents the three-body forces
\be
W^{(3)} = \frac{1}{36}\sum\limits_{123456}{\bar w}_{123456}{\psi^{\dagger}}_1{\psi^{\dagger}}_2{\psi^{\dagger}}_3\psi_6\psi_5\psi_4.
\ee
Here and in the following $\psi_1$ and $\psi^{\dagger}_1$ are fermionic field operators in some basis, whose states are completely characterized by the number indices. 
In the latter definitions we used the antisymmetrized matrix elements ${\bar v}_{1234}$ and ${\bar w}_{123456}$. Further in this work we consider  the equations of motion assuming that the Hamiltonian is confined by the two-body interaction. From the narrative it will be clear how the theory can be naturally extended to three-body and multiparticle forces. 
The fermionic fields obey the usual anticommutation relations
\bea
[\psi_1,{\psi^{\dagger}}_{1'}]_+ \equiv \psi_1{\psi^{\dagger}}_{1'}  +  {\psi^{\dagger}}_{1'}\psi_1 = \delta_{11'}, \nonumber \\
\left[ \psi_1,{\psi}_{1'} \right]_{+}  = \left[ {\psi^{\dagger}}_1,{\psi^{\dagger}}_{1'}\right]_+ = 0,
\label{anticomm}
\eea
whereas their time evolution can be described conveniently by transforming them to the Heisenberg picture:
\be
\psi(1) = e^{iHt_1}\psi_1e^{-iHt_1}, \ \ \ \ \ \ {\psi^{\dagger}}(1) = e^{iHt_1}{\psi^{\dagger}}_1e^{-iHt_1}.
\ee

Let us consider the fermionic propagator, or correlation function, in a system of $N$ interacting fermions,
which is directly linked to the observables, such as the energies of neighboring $N \pm 1$ systems and the single-particle spectroscopic strengths. $N$ is supposed to be an even integer number.
The fermionic in-medium propagator, or Green function, is defined as a correlator of two fermionic field operators :
\be
G(1,1') \equiv G_{11'}(t-t') = -i\langle T \psi(1){\psi^{\dagger}}(1') \rangle,
\label{spgf}
\ee
where $T$ is the chronological ordering operator, and the averaging $\langle ... \rangle$ is performed over the formally exact ground state of the many-body system of $N$ particles.

The basis of choice is the one
which diagonalizes the one-body (also named single-particle) part of the Hamiltonian (\ref{Hamiltonian1}): $h_{12} =  \delta_{12}\varepsilon_1$.
The convenience of using this basis will become obvious below.
The single-particle propagator (\ref{spgf}) depends explicitly on a single time difference $\tau = t-t'$, and the Fourier transform with respect to $\tau$ to the energy domain leads to its spectral (Lehmann) representation:
\bea
G_{11'}(\varepsilon) = \sum\limits_{n}\frac{\eta^{n}_{1}\eta^{n\ast}_{1'}}{\varepsilon - (E^{(N+1)}_{n} - E^{(N)}_0)+i\delta} +  \nonumber \\
+ \sum\limits_{m}\frac{\chi^{m}_{1}\chi^{ m\ast}_{1'}}{\varepsilon + (E^{(N-1)}_{m} - E^{(N)}_0)-i\delta}. 
\label{spgfspec}
\eea
It consists of terms of the simple pole character with factorized residues, that is the common feature of the propagators. The poles are located at the energies $E^{(N+1)}_{n} - E^{(N)}_0$ and $-(E^{(N-1)}_{m} - E^{(N)}_0)$ of the neighboring $(N+1)$-particle and $(N-1)$-particle systems, respectively, related to the ground state of the reference $N$-particle system.
The corresponding residues are composed of matrix elements of the field operators between the ground state $|0^{(N)}\rangle$ of the $N$-particle system and states $|n^{(N+1)} \rangle$ and $|m^{(N-1)} \rangle$ of the neighboring systems:
\be
\eta^{n}_{1} = \langle 0^{(N)}|\psi_1|n^{(N+1)} \rangle , \ \ \ \ \ \ \ \  \chi^{m}_{1} = \langle m^{(N-1)}|\psi_1|0^{(N)} \rangle .
\label{etachi}
\ee
By definition, these matrix elements give the weights of the given single-particle (single-hole) configuration on top of the ground state $|0^{(N)}\rangle$ in the $n$-th or ($m$-th) state of the $(N+1)$-particle  ($(N-1)$-particle) systems. The residues are, thereby, associated with the observable occupancies of the corresponding states.

The two most commonly used two-point two-fermion correlation functions are the particle-hole propagator, often called response function, and the particle-particle, or fermionic pair, propagator. 
The former is defined as follows:
\bea
R(12,1'2') &\equiv& R_{12,1'2'}(t-t') =   -i\langle T\psi^{\dagger}(1)\psi(2)\psi^{\dagger}(2')\psi(1')\rangle \nonumber \\ &=&
-i\langle T(\psi^{\dagger}_1\psi_2)(t)(\psi^{\dagger}_{2'}\psi_{1'})(t')\rangle,
\label{phresp}
\eea
while the latter has the form:
\bea
G(12,1'2') &\equiv& G_{12,1'2'}(t-t') = -\langle T \psi(1)\psi(2){\psi^{\dagger}}(2'){\psi^{\dagger}}(1')\rangle \nonumber \\ &=& 
-\langle T(\psi_1\psi_2)(t)(\psi^{\dagger}_{2'}\psi^{\dagger}_{1'})(t')\rangle,
\label{ppGF} 
\eea
where we imply that $t_1 = t_2 = t, t_{1'} = t_{2'} = t'$. 
The Fourier transformation of Eq. (\ref{phresp}) to the energy (frequency) domain leads to the spectral expansion
\be
R_{12,1'2'}(\omega) = \sum\limits_{\nu>0}\Bigl[ \frac{\rho^{\nu}_{21}\rho^{\nu\ast}_{2'1'}}{\omega - \omega_{\nu} + i\delta} -  \frac{\rho^{\nu\ast}_{12}\rho^{\nu}_{1'2'}}{\omega + \omega_{\nu} - i\delta}\Bigr]
\label{respspec}
\ee
which, similarly to the one for the one-fermion propagator (\ref{spgfspec}), satisfies the general requirements of locality and unitarity. The  matrix elements in the residues
\be
\rho^{\nu}_{12} = \langle 0|\psi^{\dagger}_2\psi_1|\nu \rangle 
\label{rho}
\ee
are the transition densities. They give the weights of the pure particle-hole configurations on top of the ground state $|0\rangle$ in the model  excited states $|\nu\rangle$ of the same N-particle system. The poles of the response function (\ref{respspec}) are located at the excitation energies $\omega_{\nu} = E_{\nu} - E_0$ relative to the ground state. The Fourier image of the two-time two-fermion Green function (\ref{ppGF}) reads
\be
iG_{12,1'2'}(\omega) =  \sum\limits_{\mu} \frac{\alpha^{\mu}_{21}\alpha^{\mu\ast}_{2'1'}}{\omega - \omega_{\mu}^{(++)}+i\delta} - \sum\limits_{\varkappa} \frac{\beta^{\varkappa\ast}_{12}\beta^{\varkappa}_{1'2'}}{\omega + \omega_{\varkappa}^{(--)}-i\delta},
\label{resppp}
\ee
where the residues are products of the matrix elements
\be
\alpha_{12}^{\mu} = \langle 0^{(N)} | \psi_2\psi_1|\mu^{(N+2)} \rangle , \ \ \ \ \ \ \beta_{12}^{\varkappa} = \langle 0^{(N)} | \psi^{\dagger}_2\psi^{\dagger}_1|\varkappa^{(N-2)} \rangle
\label{alphabeta}
\ee
and the poles $\omega_{\mu}^{(++)} = E_{\mu}^{(N+2)} - E_0^{(N)}$ and  $\omega_{\varkappa}^{(--)} = E_{\varkappa}^{(N-2)} - E_0^{(N)}$ are formally exact states of the $(N+2)$- and $(N-2)$-particle systems, respectively.

Obviously, Eqs. (\ref{spgfspec},\ref{respspec},\ref{resppp}) are model independent and valid for any physical approximations to the many-body states $|n\rangle, |m\rangle$, $|\nu\rangle$, $|\mu\rangle$, and $|\varkappa\rangle$. In Eqs. (\ref{spgfspec},\ref{respspec},\ref{resppp})  the sums are formally complete, i.e., run over the discrete spectra and engage the corresponding integrals over the continuum states.

\section{One-fermion propagator: the equation of motion (EOM)}
\label{EOM1}
\subsection{Normal phase}
\label{normal}
The time evolution of the fermionic propagator (\ref{spgf}) can be traced by taking its time derivatives. The differentiation with respect to $t$ leads to
\bea
\partial_t G_{11'}(t-t') = -i\delta(t-t')\langle [\psi_1(t),{\psi^{\dagger}}_{1'}(t')]_+\rangle + \nonumber \\
+ \langle T[H,\psi_1](t){\psi^{\dagger}}_{1'}(t')\rangle, \nonumber\\ 
\label{dtG}                           
\eea
where 
$[H,\psi_1](t) = e^{iHt}[H,\psi_1]e^{-iHt}.$
Evaluating the commutator and isolating the terms with $G_{11'}(t-t')$, one obtains the equation:
\be
(i\partial_t - \varepsilon_1)G_{11'}(t-t') = \delta_{11'}\delta(t-t') + i\langle T[V,\psi_1](t){\psi^{\dagger}}_{1'}(t')\rangle,
\label{spEOM}
\ee
which is commonly referred to as the first EOM, or EOM1. The second EOM is generated by the differentiation of the last term on the right hand side of Eq. (\ref{spEOM}),
\be
R_{11'}(t-t') =  i\langle T[V,\psi_1](t){\psi^{\dagger}}_{1'}(t')\rangle, 
\ee
with respect to $t'$:
\bea
R_{11'}(t-t')\overleftarrow{\partial_{t'}} 
&=& -i\delta(t-t') \langle \bigl[[V,\psi_1](t),{\psi^{\dagger}}_{1'}(t')\bigr]_+\rangle - \nonumber \\ 
&-& \langle T[V,\psi_1](t)[H,{\psi^{\dagger}}_{1'}](t')\rangle,
\eea
which gives the second EOM, or EOM2:
\bea
R_{11'}(t-t')(-i\overleftarrow{\partial_{t'}}  - \varepsilon_{1'}) &=& -\delta(t-t')\langle \bigl[ [V,\psi_1](t),{\psi^{\dagger}}_{1'}(t')\bigr]_+\rangle\nonumber \\
&+& i\langle T [V,\psi_1](t)[V,{\psi^{\dagger}}_{1'}](t')\rangle.
\label{EOMR}
\eea
Acting on the EOM1 (\ref{spEOM}) by the operator $(-i\overleftarrow{\partial_{t'}}  - \varepsilon_{1'})$ and performing the Fourier transformation to the energy domain with respect to the time difference $t-t'$ yield:
\be
G_{11'}(\omega) 
= G^{0}_{11'}(\omega) + 
\sum\limits_{22'}G^{0}_{12}(\omega)T_{22'}(\omega)G^{0}_{2'1'}(\omega)
\label{spEOM3}
\ee
with the free, or uncorrelated, fermionic propagator $G^{0}_{11'}(\omega) = \delta_{11'}/(\omega - \varepsilon_1)$
and the interaction kernel (or the one-body $T$-matrix):
\bea
T_{11'}(t-t') &=& T^{0}_{11'}(t-t') + T^{r}_{11'}(t-t'), \nonumber\\
T^{0}_{11'}(t-t') &=& -\delta(t-t')\langle \bigl[ [V,\psi_1](t),{\psi^{\dagger}}_{1'}(t')\bigr]_+\rangle, \nonumber \\ 
T^{r}_{11'}(t-t') &=&  i\langle T [V,\psi_1](t)[V,{\psi^{\dagger}}_{1'}](t')\rangle.
\label{Toperator}
\eea
The superscript "0" is associated with the static parts of the interaction kernels and "r" with their dynamical, or time-dependent, parts, which take care of retardation effects. 
The EOM (\ref{spEOM3}) can be written in the operator form as:
\be
G(\omega) = G^{0}(\omega) + G^{0}(\omega)T(\omega)G^{0}(\omega).
\label{spEOM4}
\ee
To transform it to the Dyson equation, one introduces the irreducible part of the $T$-matrix, the self-energy, such as $\Sigma = T^{irr}$. The irreducibility here is taken with respect to the uncorrelated one-fermion propagator $G^{0}$, so that the self-energy and the $T$-matrix are related as follows:
\be
T(\omega) = \Sigma(\omega) + \Sigma(\omega) G^{0}(\omega)T(\omega).
\label{DysonT}
\ee
Eliminating the $T$-matrix, one arrives at the Dyson equation for fermionic propagator:
\be
G(\omega) = G^{0}(\omega) + G^{0}(\omega)\Sigma(\omega) G(\omega).
\label{Dyson}
\ee
The self-energy consists of the instantaneous mean-field part $\Sigma^{0}$ and the energy(time)-dependent, or  dynamical, part  $\Sigma^{r}(\omega)$, as it follows from the expression for the $T$-matrix (\ref{Toperator}):
\be
\Sigma_{11'}(\omega) = \Sigma_{11'}^{0} + \Sigma_{11'}^{r}(\omega).
\label{Somega}
\ee

The explicit evaluations of the commutators in Eq. (\ref{Toperator}) give for the static (instantaneous) term:
\be
 \Sigma^{0}_{11'} = -\langle[[V,\psi_1],{\psi^{\dagger}}_{1'}]_+\rangle = \sum\limits_{il}{\bar v}_{1i1'l}\rho_{li}, 
 \label{MF}
\ee
where  $\rho_{li} = \langle{\psi^{\dagger}}_i\psi_l\rangle$  is the ground-state one-body density.
To obtain the dynamical part $\Sigma^{r}(\omega)$ of the mass operator, which comprises all retardation effects induced by the nuclear medium, we first compute its reducible counterpart $T^{r}_{11'}(t-t')$:
\bea
T^{r}_{11'}(t&-&t') = -\frac{i}{4} \sum\limits_{2'3'4'}\sum\limits_{234}{\bar v}_{1234}\times \nonumber\\
&\times&\langle T \psi^{\dagger}(2)\psi(4)\psi(3)\psi^{\dagger}(3')\psi^{\dagger}(4')\psi(2')\rangle
{\bar v}_{4'3'2'1'},  \nonumber\\
\label{Tr}
\eea
where we assume that $t_{2}=t_{3}=t_{4}=t$ and $t_{2'}=t_{3'}=t_{4'}=t'$, as constrained by the instantaneous interaction. 
Here it becomes clear that, although the EOM for the fermionic propagator $G(\omega)$ (\ref{Dyson}) is formally  a closed equation with respect to $G(\omega)$, its interaction kernel is defined by the three-fermion Green function.  The EOM for the three-body propagator, however, generates even higher-rank propagators, which makes the exact solution of the many-body problem intractable. Instead of reducing  the problem to the perturbative expansions in powers of the interaction, we proceed with the cluster decomposition of the kernel, which allows for truncation of the many-body problem at the two-body level. More precisely  \cite{Martin1959,VinhMau1969,Mau1976,RingSchuck1980}:
 \begin{eqnarray}
 &i&\langle T \psi^{\dagger}(2)\psi(4)\psi(3)\psi^{\dagger}(3')\psi^{\dagger}(4')\psi(2')\rangle \equiv 
 G(432^{\prime},23^{\prime}4^{\prime}) \approx\nonumber \\
 &\approx& G(4,4')G(32',23') + G(3,3')G(42',24') +  \nonumber \\ &+& G(2',2)G(43,3'4') + G(4,2)G(32',3'4') +    \nonumber \\ 
 &+& G(2',4')G(43,23') - G(3,2)G(42',3'4')  - \nonumber \\ 
 &-& G(2',3')G(43,24') - G(4,3')G(32',24')  - \nonumber \\ &-& G(3,4') G(42',23') - 2G^{(0)}(432^{\prime},23^{\prime}4^{\prime}),\nonumber \\
 \label{G3}
 \end{eqnarray}
 where
 \begin{eqnarray}
 &G&^{(0)}(432^{\prime},23^{\prime}4^{\prime}) = \nonumber \\ &=& - G(4,4')G(3,3')G(2',2) + G(4,3')G(3,4')G(2',2) + \nonumber \\ 
 &+& G(4,2)G(3,3')G(2',4')  + G(4,4')G(3,2)G(2',3') - \nonumber \\ 
 &-&G(4,2)G(3,4')G(2',3') - G(4,3')G(3,2)G(2',4') 
 \label{G03}
 \end{eqnarray}
contains all uncorrelated three-body contributions. The approximation of Eq. (\ref{G3}) neglects only the fully correlated three-body terms.
Dropping the reducible contributions (terms containing equal-times one-fermion propagators) in Eq. (\ref{G3}), we obtain:
\begin{eqnarray}
  \Sigma^{r}_{11'}(t-t') =-\sum\limits_{2342'3'4'}{\bar v}_{1234}\Bigl(\frac{1}{4}G(2',2){G}^{(pp)}(43,3'4') \nonumber \\ - iG(3,3'){R}^{(ph)}(24,2'4') + G(3,3')G(2',2)G(4,4')\Bigr){\bar v}_{4'3'2'1'}  
\nonumber \\
=  \Sigma^{r(pp)}_{11'}(t-t') +  \Sigma^{r(ph)}_{11'}(t-t') +  \Sigma^{r(0)}_{11'}(t-t'),\nonumber \\
 \label{SEirr2}
 \end{eqnarray}
where the  particle-particle ($pp$) and particle-hole ($ph$) character of the correlation functions defined by Eqs. (\ref{phresp},\ref{ppGF}) are marked explicitly.

\begin{figure}
\begin{center}
\includegraphics[scale=0.52]{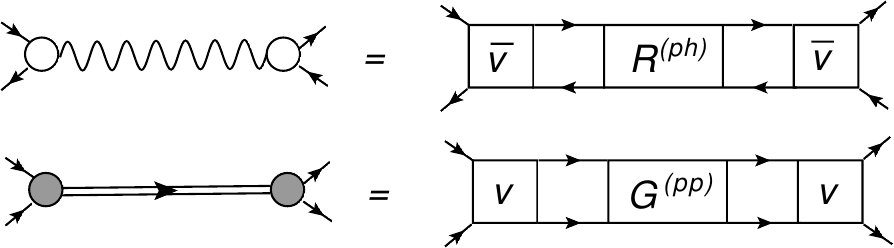}
\end{center}
\caption{The mapping of the phonon vertices (empty and filled circles) and propagators (wavy lines and double lines) onto the bare interaction (squares, antisymmetrized $\bar v$ and plain $v$) and two-fermion correlation functions (rectangular blocks $R^{(ph)}$ and $G^{(pp)}$) in diagrammatic form. Lines with arrows stand for fermionic particles (right arrows) and holes (left arrows). Top: normal (particle-hole) phonon, bottom: pairing (particle-particle) phonon,  as introduced in Eqs. (\ref{mappingph},\ref{mappingpp}), respectively.}
\label{PVCmap}%
\end{figure}

The self-energy of Eq. (\ref{SEirr2})  serves as foundation for microscopic approaches to the single-particle self-energy, which refer to the phenomenon of particle-vibration coupling, or PVC. To show this explicitly, one can identify the correlation functions $R^{(ph)}$ and $G^{(pp)}$ contracted with the interaction matrix elements with the phonon propagators and coupling vertices. This mapping is displayed diagrammatically in Fig. \ref{PVCmap}.
At this point it is convenient to work with the Fourier image of $\Sigma^{r}_{11'}(t-t')$ in the energy domain: 
\be
\Sigma^{r}_{11'}(\omega) = \int\limits_{-\infty}^{\infty} d\tau e^{i\omega\tau} \Sigma^{r}_{11'}(\tau).
\label{SEomega}
\ee
The transformation of the first term of Eq. (\ref{SEirr2}) reads:
\bea
\Sigma^{r(pp)}_{11'}(\omega) = \sum\limits_{22'} \Bigl[ \sum\limits_{\mu m} \frac{\chi_2^{m\ast} \gamma_{12}^{\mu(+)}\gamma_{1'2'}^{\mu(+)\ast}\chi_{2'}^{m}}{\omega - \omega_{\mu}^{(++)} - \varepsilon_m^{(-)} + i\delta} + \nonumber\\
+ \sum\limits_{\varkappa n}\frac{\eta_2^{n\ast}{\gamma}_{21}^{\varkappa(-)\ast}{\gamma}_{2'1'}^{\varkappa(-)}\eta_{2'}^n}{\omega + \omega_{\varkappa}^{(--)} + \varepsilon_n^{(+)} - i\delta} \Bigr] .
\label{FISrpp}
\eea
Here the single-particle energies in the neighboring $(N+1)$-particle system are denoted as $\varepsilon_n^{(+)} = E^{(N+1)}_{n} - E^{(N)}_0$ and those in the neighboring $(N-1)$-particle system as $\varepsilon_m^{(-)} = E^{(N-1)}_{m} - E^{(N)}_0$. The pairing phonon vertex functions are then defined as follows:
\be
\gamma^{\mu(+)}_{12} = \sum\limits_{34} v_{1234}\alpha_{34}^{\mu}, \ \ \ \ \ \ \gamma_{12}^{\varkappa(-)} = \sum\limits_{34}\beta_{34}^{\varkappa}v_{3412}. 
\label{vert_pp}
\ee
Sometimes it is convenient to ntroduce the pairing interaction amplitude $\Gamma^{pp}_{12,1'2'}(\omega) $,
\bea
i\Gamma^{pp}_{12,1'2'}(\omega) = i\sum\limits_{343'4'}{v}_{1234}G^{(pp)}_{43,3'4'}(\omega){v}_{4'3'2'1'} = \nonumber \\
= \sum\limits_{\mu,\sigma=\pm1} \gamma^{\mu(\sigma)}_{12}\Delta^{(\sigma)}_{\mu}(\omega)\gamma^{\mu(\sigma)\ast}_{1'2'}
\label{mappingpp}
\eea
as the contraction of these vertices with the pairing phonon propagator:
\be
\Delta^{(\sigma)}_{\mu}(\omega) = \frac{\sigma}{\omega - \sigma(\omega_{\mu}^{(\sigma\sigma)} - i\delta)}.
\ee
Then, Eq. (\ref{FISrpp}) can be alternatively obtained by the following convolution:
\be
\Sigma^{r(pp)}_{11'}(\omega) = i\sum\limits_{22'}\int\limits_{-\infty}^{\infty}\frac{d\varepsilon}{2\pi i} \Gamma^{pp}_{12,1'2'}(\omega + \varepsilon)G_{2'2}(\varepsilon).
\ee
\begin{figure*}
\begin{center}
\includegraphics*[scale=0.75]{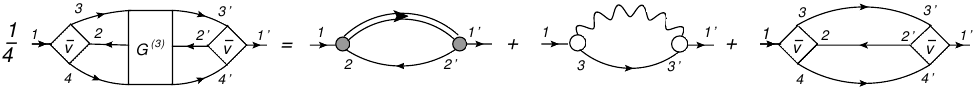}
\end{center}
\caption{The dynamical kernel $\Sigma^{r}$ of Eq. (\ref{SEirr2}) in terms of the particle-vibration coupling. Same conventions as in Fig. \ref{PVCmap} apply, while the rectangular block $G^{(3)}$ stands for the three-fermion propagator of Eq. (\ref{G3}).}
\label{SEdyn}%
\end{figure*}
The Fourier image of the second term of Eq. (\ref{SEirr2}) can be treated similarly:
\bea
\Sigma^{r(ph)}_{11'}(\omega) = \sum\limits_{33'} \Bigl[ 
\sum\limits_{\nu n}\frac{\eta_3^{n}{g}_{13}^{\nu}{g}_{1'3'}^{\nu\ast}\eta_{3'}^{n\ast}}{\omega - \omega_{\nu} - \varepsilon_n^{(+)} + i\delta} +
\nonumber\\ +
\sum\limits_{\nu m} \frac{\chi_3^{m} g_{31}^{\nu\ast}g_{3'1'}^{\nu}\chi_{3'}^{m\ast}}{\omega + \omega_{\nu} + \varepsilon_m^{(-)} - i\delta} 
\Bigr],
\eea
if the following mapping is performed:
\bea
\Gamma^{ph}_{13',1'3} =  \sum\limits_{242'4'}{\bar v}_{1234}R^{(ph)}_{24,2'4'}(\omega){\bar v}_{4'3'2'1'} = \nonumber \\ = 
\sum\limits_{\nu,\sigma=\pm1} g^{{\nu}(\sigma)}_{13}D^{(\sigma)}_{\nu}(\omega)g^{\nu(\sigma)\ast}_{1'3'},
\label{mappingph}
\eea
where we introduced the phonon vertices $g^{\nu}$ and propagators $D_{\nu}(\omega)$:
\bea
g^{\nu(\sigma)}_{13} = \delta_{\sigma,+1}g^{\nu}_{13} + \delta_{\sigma,-1}g^{\nu\ast}_{31}, \ \ \ \ 
g^{\nu}_{13} = \sum\limits_{24}{\bar v}_{1234}\rho^{\nu}_{42}, 
\label{vert_ph} \nonumber\\
\\
D_{\nu}^{(\sigma)}(\omega) = \frac{\sigma}{\omega - \sigma(\omega_{\nu} - i\delta)}, \ \ \ \
\omega_{\nu} = E_{\nu} - E_0. \nonumber \\
\label{gDPVCph}
\eea 
The corresponding integral expression is:
\be
\Sigma^{r(ph)}_{11'}(\omega) = -\sum\limits_{33'}\int\limits_{-\infty}^{\infty}\frac{d\varepsilon}{2\pi i} \Gamma^{ph}_{13',1'3}(\omega - \varepsilon)G_{33'}(\varepsilon).
\label{FISphe}
\ee
The new index $\sigma = \pm1$ in Eqs. (\ref{FISrpp}-\ref{FISphe}) was introduced to indicate the forward ("particle") and backward ("hole") components of the phonon propagators, that also affects the vertices. The spectral representations (\ref{spgfspec},\ref{resppp}) along with the definitions (\ref{spgf},\ref{ppGF},\ref{phresp}) were applied to implement the mapping. 
The last term of the self-energy (\ref{SEirr2}) with the uncorrelated single-particle Green functions transforms to the energy domain as follows:
\bea
&\Sigma&^{r(0)}_{11'}(\omega) = -\sum\limits_{2342'3'4'} {\bar v}_{1234}\times \nonumber \\ &\times&\Bigl[\sum\limits_{mn'n''} \frac{\chi_{2'}^{m}\chi_2^{m\ast}\eta_3^{n'}\eta_{3'}^{n'\ast}\eta_4^{n''}\eta_{4'}^{n''\ast}}
{\omega - \varepsilon_{n'}^{(+)} - \varepsilon_{n''}^{(+)} - \varepsilon_{m}^{(-)} + i\delta} \nonumber \\
&+& \sum\limits_{nm'm''} \frac{\eta_{2'}^{n}\eta_{2}^{n\ast}\chi_3^{m'}\chi_{3'}^{m'\ast}\chi_4^{m''}\chi_{4'}^{m''\ast}}
{\omega + \varepsilon_{n}^{(+)} + \varepsilon_{m'}^{(-)} + \varepsilon_{m''}^{(-)} - i\delta} \Bigr] {\bar v}_{4'3'2'1'} = \nonumber \\
&=&-\sum\limits_{2342'3'4'}{\bar v}_{1234} {\tilde G}^{(3)0}_{432',23'4'}(\omega){\bar v}_{4'3'2'1'},
\\
&{\tilde G}&^{(3)0}_{432',23'4'} (\omega) = \nonumber \\
&=& -\int\limits_{-\infty}^{\infty}\frac{d\varepsilon d\varepsilon'}{(2\pi i)^2}  G_{44'}(\omega+\varepsilon'-\varepsilon)G_{33'}(\varepsilon)G_{2'2}(\varepsilon').
\eea

The full dynamical part of the fermionic self-energy (\ref{SEirr2}) is shown in Fig. \ref{SEdyn} in the diagrammatic form.  The particle-vibration coupling vertices are denoted by the circles (empty for the normal phonon and filled for the pairing ones), the propagator of the normal phonons is associated with the wavy line and that of the pairing phonons is given by the double line with an arrow.  Note that the signs in front of the diagrams depend on the diagrammatic conventions, for instance, they may not respect the Feynman's convention. For instance, the last uncorrelated term is often shown with the "-" sign in the literature, and the phonon vertices may include the multiplier "$i$".

The first two diagrams on the right hand side in Fig. \ref{SEdyn} are the topologically similar one-loop diagrams, which are analogous to the electron self-energy corrections in quantum electrodynamics 
(QED), 
where electron emits and reabsorbs a photon. In the nucleonic self-energy of quantum hadrodynamics (QHD) a single nucleon emits and reabsorbs  mesons with various quantum numbers. 
In the present context, the first two diagrams of Fig. \ref{SEdyn}  represent the effects of a strongly correlated medium, where a single fermion emits and reabsorbs phonons of the particle-particle and the particle-hole nature. In this way, the phonons emerge as effective mediators of interaction, additional to the original bare interaction between two fermions, which is stipulated by the correlated medium. This fact can be expresses by introducing an effective Hamiltonian containing the explicit phonon degrees of freedom, that is often used in phenomenological approaches. Fig. \ref{PVCmap} illustrates diagrammatically the mappings introduced by Eqs. (\ref{mappingpp},\ref{mappingph}). This mapping is a key point for the present discussion as it emphasizes the essentially non-perturbative character of the approach and explains the underlying mechanism of the induced in-medium interaction. It also clarifies the difference between the present case and QED or QHD, where the fermionic and bosonic degrees of freedom are completely independent. In contrast, here the emergent composite bosons are formed by correlated fermionic pairs. Importantly, their couplings are not the effective parameters of the theory, but can be calculated consistently from the underlying fermion-fermion bare interaction.

The dynamical self-energy $\Sigma^{r}$ in the form of Eq. (\ref{SEirr2}) and Fig. \ref{SEdyn} helps to relate the approach with the phonon exchange to the lowest-order perturbation theory and, thus, to assess the role of complex correlations. In the case of weak coupling the uncorrelated term(s) play the leading role and the phonon-exchange interaction can be neglected. In the strong-coupling regime, in contrast, the phonon coupling dominates over the lowest-order uncorrelated term (third term on the right hand side of Fig. \ref{SEdyn}), so that the lowest-order approach does not produce the leading contribution. However, the situation may differ in the frameworks based on the effective interactions. The latter are typically obtained by fitting the bulk nuclear properties, such as their masses and radii, on the Hartree or Hartree-Fock level assuming that fermionic self-energy contains only the static part (\ref{MF}) dependent on the one-body density. This density is supposed to be implicitly coupled to correlations in the dynamical part of the self-energy (\ref{SEirr2}). This becomes obvious if one notices that the one-fermion density is the equal-times limit of the one-fermion propagator
\be
\rho_{12} = -i\lim\limits_{t_2\to t_1+0} G(1,2),
\ee
given by the full solution of Eq. (\ref{Dyson}), which absorbs correlations also from the dynamical part. This fact is typically expressed in terms of the density dependencies of the effective interactions, however, these dependencies do not follow from a detailed analysis of Eq. (\ref{Dyson}) with the complete kernel $\Sigma(\omega)$.
The existing versions of the PVC model, which take into account the dynamical self-energy on top of the effective interactions inevitably imply an additional procedure to remove the double counting of PVC. The general reasoning for the appearance of this double counting is that the PVC is already contained in the parameters of the phenomenological mean field \cite{Tselyaev2013}. An explicit subtraction of the dynamical PVC contribution taken in the static limit from the effective interaction turned out to be an elegant way of avoiding such a double counting.  The subtraction method is widely applied in calculations of two-body Green functions, in particular, the particle-hole response \cite{Tselyaev2013,LitvinovaTselyaev2007,LitvinovaRingTselyaev2007,LitvinovaRingTselyaev2008,LitvinovaSchuck2019}. For the case of the one-body propagator this method has not been adopted yet.

Computation of the fermionic self-energy with PVC requires the knowledge about the two-fermion propagators $R^{(ph)}$ and $G^{(pp)}$ or, equivalently, the phonon vertices and propagators. They can be found by solving the EOM's for these correlation functions together with Eq. (\ref{Dyson}). The analyses of the corresponding EOM's can be found in Refs. \cite{DukelskyRoepkeSchuck1998,Olevano2018,LitvinovaSchuck2019,LitvinovaSchuck2020}.

\subsection{Superfluid phase}
\label{superfluid}

The superfluid phase is characterized by the pronounced formation of Cooper pairs and, thus, an enhanced role of the pairing phonons. While in calculations for normal systems the PVC approach to the self-energy usually neglects the term with the pairing phonons because of its relatively low importance, the situation may be different for superfluid systems. Within the PVC approach discussed above the pairing interaction is fully dynamical and mediated by the pairing phonons emerging naturally in the one-fermion self-energy. In the traditional frameworks based on effective interactions, however, the pairing is included in the static approximations like the Bardeen-Cooper-Schrieffer or the Bogoliubov ones. On this level of description, the corresponding Green function technique is the Gor'kov Green functions, that can be obtained from the EOM1 if the two-body correlations are neglected (see Appendix for details).

For the ab-initio approaches with dynamical self-energies, however, the idea of introducing the anomalous Green functions  can be also very fruitful in the case of using the basis which already includes the superfluidity effects in some static approximation. Those can be, for instance, the Hartree-(Fock)-BCS (HF-BCS) or the Hartree-(Fock)-Bogoliubov (HFB) bases. Since the states in these bases are of the quasiparticle character, the space of the single-particle variables is doubled by the Bogoliubov transformation of the fermionic field operators: 
\bea
\psi_1 = U_{1\mu}\alpha_{\mu} + V^{\ast}_{1\mu}\alpha^{\dagger}_{\mu}\nonumber\\
\psi^{\dagger}_1 = V_{1\mu}\alpha_{\mu} + U^{\ast}_{1\mu}\alpha^{\dagger}_{\mu},
\label{Btrans}
\eea
where summation is implied over the repeated index $\mu$, or, in the operator form:
\bea
\left( \begin{array}{c} \psi \\ \psi^{\dagger} \end{array} \right) = \cal{W} \left( \begin{array}{c} \alpha \\ \alpha^{\dagger} \end{array} \right),
\eea
where
\bea
\cal{W} = \left( \begin{array}{cc} U & V^{\ast} \\ V & U^{\ast} \end{array} \right) \ \ \ \ \ \  \cal{W}^{\dagger} = \left( \begin{array}{cc} U^{\dagger} & V^{\dagger} \\ V^T & U^T \end{array} \right). 
\eea
In Eq. (\ref{Btrans}) and henceforth the Greek indices will be used to denote fermionic states in the HFB basis, while the number indices and the Roman indices introduced below will be reserved for the single-particle mean-field basis states.
The transformation $\cal W$ is unitary, and the quasiparticle operators $\alpha$ and $\alpha^{\dagger}$ form the same anticommutator algebra as the particle operators $\psi$ and $\psi^{\dagger}$, so that the matrices $U$ and $V$ satisfy:
\bea
U^{\dagger}U + V^{\dagger}V = \mathbb{1}\ \ \ \ \ \ UU^{\dagger} + V^{\ast}V^{T} = \mathbb{1}\nonumber\\
U^TV + V^TU = 0\ \ \ \ \ \  UV^{\dagger} + V^{\ast}U^{T} = 0 .
\label{UV}
\eea

To use consistently the HFB basis, which will be referred to as {\it quasiparticle} basis, for the description of the fermionic propagator, the latter should be also extended. This can be done with the aid of the generalized field operators $\Psi$ and $\Psi^{\dagger}$
\be
\Psi_1(t_1) = \left( \begin{array}{c} \psi_1(t_1) \\ \psi_1^{\dagger}(t_1) \end{array} \right), \ \ \ \ \ \ \ \ \ 
\Psi^{\dagger}_1(t_1) = \Bigl( \psi_1^{\dagger}(t_1) \ \ \ \ \ \psi_1(t_1) \Bigr)
\label{psicol}
\ee
to the doubled single-particle space:
\bea
{\hat G}_{12}(t-t') = -i\langle T\Psi_1(t)\Psi^{\dagger}_2(t')\rangle = \nonumber\\
= -i\theta(t-t')\left( \begin{array}{cc} \langle \psi_1(t)\psi^{\dagger}_2(t')\rangle &  \langle \psi_1(t)\psi_2(t')\rangle \\
 \langle \psi^{\dagger}_1(t)\psi^{\dagger}_2(t')\rangle &  \langle \psi^{\dagger}_1(t)\psi_2(t')\rangle
\end{array} \right)  + \nonumber \\
+ i\theta(t'-t)\left( \begin{array}{cc} \langle \psi^{\dagger}_2(t')\psi_1(t)\rangle &  \langle \psi_2(t')\psi_1(t)\rangle \\
 \langle \psi^{\dagger}_2(t')\psi^{\dagger}_1(t)\rangle &  \langle \psi_2(t')\psi^{\dagger}_1(t)\rangle
\end{array} \right). 
\label{GG}
\eea
Notice here that the equal-times limit of this propagator is the Valatin density matrix \cite{Valatin1961}
\bea
{\cal R}_{12} =  -i\lim\limits_{t' \to t+0} G_{12}(t-t') = \nonumber \\ 
= \left( \begin{array}{cc} \langle \psi^{\dagger}_2\psi_1\rangle &  \langle \psi_2\psi_1\rangle \\
 \langle \psi^{\dagger}_2\psi^{\dagger}_1\rangle &  \langle \psi_2\psi^{\dagger}_1\rangle 
\end{array} \right)
\equiv \left( \begin{array}{cc} \rho_{12} & \varkappa_{12}\\ -\varkappa_{12}^{\ast} & 1-\rho^{\ast}_{12} \end{array}\right)
\label{Valatin}
\eea
commonly adopted in the many-body theory \cite{RingSchuck1980}.
The Fourier transform of the generalized propagator (\ref{GG}) reads:
\bea
{\hat G}_{12}(\varepsilon) = \sum\limits_{n}\frac{\langle 0|\Psi_1|n\rangle \langle n|\Psi^{\dagger}_2|0\rangle}{\varepsilon - (E^{(N+1)}_{n} - E^{(N)}_0)+i\delta} +  \nonumber \\
+ \sum\limits_{m}\frac{\langle 0|\Psi^{\dagger}_2|m\rangle \langle m|\Psi_1|0\rangle}{\varepsilon + (E^{(N-1)}_{m} - E^{(N)}_0)-i\delta}. 
\label{spgfspec_gen}
\eea
The basis-dependent quantities here are the matrix elements in the residues. Their transformation to the quasiparticle basis can be performed with the aid of Eqs. (\ref{Btrans},\ref{psicol}):
\bea
\langle 0|\Psi_1|n\rangle  &=& \left( \begin{array}{c} \langle 0|\psi_1|n\rangle \\ \langle 0|\psi_1^{\dagger} |n\rangle \end{array} \right) =
\left( \begin{array}{c} \langle 0|U_{1\mu}\alpha_{\mu} + V^{\ast}_{1\mu}\alpha^{\dagger}_{\mu}|n\rangle \\ 
\langle 0|V_{1\mu}\alpha_{\mu} + U^{\ast}_{1\mu}\alpha^{\dagger}_{\mu} |n\rangle \end{array} \right)\nonumber\\
\langle n|\Psi^{\dagger}_2|0\rangle &=& \Bigl( \langle n|\psi_2^{\dagger}|0\rangle \ \ \ \ \  \langle n|\psi_2|0\rangle \Bigr) =\nonumber\\
= &\Bigl(& \langle n|V_{2\mu}\alpha_{\mu} + U^{\ast}_{2\mu}\alpha^{\dagger}_{\mu}|0\rangle \ \ \ \ \  \langle n|U_{2\mu}\alpha_{\mu} + V^{\ast}_{2\mu}\alpha^{\dagger}_{\mu}|0\rangle \Bigr)\nonumber\\
\langle 0|\Psi^{\dagger}_2|m\rangle &=& \Bigl( \langle 0|\psi_2^{\dagger}|m\rangle \ \ \ \ \  \langle 0|\psi_2|m\rangle \Bigr) =\nonumber\\
= &\Bigl(& \langle 0|V_{2\mu}\alpha_{\mu} + U^{\ast}_{2\mu}\alpha^{\dagger}_{\mu}|m\rangle \ \ \ \ \  \langle 0|U_{2\mu}\alpha_{\mu} + V^{\ast}_{2\mu}\alpha^{\dagger}_{\mu}|m\rangle \Bigr)\nonumber\\
\langle m|\Psi_1|0\rangle  &=& \left( \begin{array}{c} \langle m|\psi_1|0\rangle \\ \langle m|\psi_1^{\dagger} |0\rangle \end{array} \right) =
\left( \begin{array}{c} \langle m|U_{1\mu}\alpha_{\mu} + V^{\ast}_{1\mu}\alpha^{\dagger}_{\mu}|0\rangle \\ 
\langle m|V_{1\mu}\alpha_{\mu} + U^{\ast}_{1\mu}\alpha^{\dagger}_{\mu} |0\rangle \end{array} \right).\nonumber\\
\label{residues}
\eea
If the ground state $|0\rangle$ of the many-body system is the quasiparticle vacuum, i.e., $\alpha|0\rangle = 0$, Eqs. (\ref{residues}) further reduce to:
\bea
\langle 0|\Psi_1|n\rangle  &=& \left( \begin{array}{c} U_{1\mu} \\ V_{1\mu} \end{array}\right)\langle 0|\alpha_{\mu}|n\rangle\nonumber\\
\langle n|\Psi^{\dagger}_2|0\rangle &=& \Bigl(U^{\ast}_{2\mu} \ \ \ \ \  V^{\ast}_{2\mu} \Bigr) \langle n|\alpha^{\dagger}_{\mu} |0\rangle\nonumber\\
\langle 0|\Psi^{\dagger}_2|m\rangle &=& \Bigl(V_{2\mu} \ \ \ \ \  U_{2\mu} \Bigr) \langle 0|\alpha_{\mu} |m\rangle\nonumber\\
\langle m|\Psi_1|0\rangle  &=& \left( \begin{array}{c} V^{\ast}_{1\mu} \\ U^{\ast}_{1\mu} \end{array}\right)\langle m|\alpha^{\dagger}_{\mu}|0\rangle
\eea
and, thus, the propagator takes the form:
\bea
{\hat G}_{12}(\varepsilon) &=&  \sum\limits_{n\mu\nu}\left( \begin{array}{c} U_{1\mu} \\ V_{1\mu} \end{array}\right)\Bigl(U^{\ast}_{2\nu} \ \ \  V^{\ast}_{2\nu} \Bigr)
\frac{\langle 0|\alpha_{\mu}|n\rangle\langle n|\alpha^{\dagger}_{\nu}|0\rangle}{\varepsilon - (E_n - E_0) + i\delta} + \nonumber\\
&+& \sum\limits_{m\mu\nu}\left( \begin{array}{c} V^{\ast}_{1\nu} \\ U^{\ast}_{1\nu} \end{array}\right)\Bigl(V_{2\mu} \ \ \  U_{2\mu} \Bigr)
\frac{\langle 0|\alpha_{\mu}|m\rangle\langle m|\alpha^{\dagger}_{\nu}|0\rangle}{\varepsilon + (E_m - E_0) - i\delta},\nonumber\\
\label{GG1}
\eea
where we omitted the superscripts $(N)$ and $(N\pm1)$ at the energies of the intermediate states, as the particle number conservation is relaxed in the present approach, and put back the explicit summations over the quasiparticle indices $\mu$ and $\nu$.
Furthermore, if the intermediate states are of the one-quasiparticle character, i.e., $|m\rangle = \alpha^{\dagger}_m|0\rangle$, Eq. (\ref{GG1}) determines the mean-field propagator as
\bea
{\hat {\tilde G}}_{12}(\varepsilon) &=&  \sum\limits_{\mu}\left( \begin{array}{c} U_{1\mu} \\ V_{1\mu} \end{array}\right)\Bigl(U^{\ast}_{2\mu} \ \ \  V^{\ast}_{2\mu} \Bigr)
\frac{\langle 0|\alpha_{\mu}|\mu\rangle\langle \mu|\alpha^{\dagger}_{\mu}|0\rangle}{\varepsilon - (E_{\mu} - E_0) + i\delta} + \nonumber\\
&+& \sum\limits_{\nu}\left( \begin{array}{c} V^{\ast}_{1\nu} \\ U^{\ast}_{1\nu} \end{array}\right)\Bigl(V_{2\nu} \ \ \  U_{2\nu} \Bigr)
\frac{\langle 0|\alpha_{\nu}|\nu\rangle\langle \nu|\alpha^{\dagger}_{\nu}|0\rangle}{\varepsilon + (E_{\nu} - E_0) - i\delta} \nonumber\\
\label{GG2}
\eea
or, equivalently,
\bea
{\hat {\tilde G}}_{12}(\varepsilon) &=&  \sum\limits_{\mu}\left( \begin{array}{cc} U_{1\mu}U^{\dagger}_{\mu 2} &  U_{1\mu}V^{\dagger}_{\mu 2} \\ 
V_{1\mu}U^{\dagger}_{\mu 2} & V_{1\mu}V^{\dagger}_{\mu 2}\end{array}\right)
\frac{1}{\varepsilon - (E_{\mu} - E_0) + i\delta} + \nonumber\\
&+& \sum\limits_{\nu} \left( \begin{array}{cc} V_{1\nu}V^{\dagger}_{\nu 2} &  V_{1\nu}U^{\dagger}_{\nu 2} \\ 
U_{1\nu}V^{\dagger}_{\nu 2} & U_{1\nu}U^{\dagger}_{\nu 2}\end{array}\right)^{\ast}
\frac{1}{\varepsilon + (E_{\nu} - E_0) - i\delta}.\nonumber\\
\label{GG3}
\eea
The matrix forms of the propagators of Eqs. (\ref{GG1}-\ref{GG3}) can be directly related to the Gor'kov-type Green functions:
\bea
{\hat G}_{12}(\varepsilon) \equiv \left( \begin{array}{cc} G^{(11)}_{12}(\varepsilon)  & G^{(12)}_{12}(\varepsilon) \\ 
G^{(21)}_{12}(\varepsilon) & G^{(22)}_{12}(\varepsilon)\end{array}\right) \equiv 
\left( \begin{array}{cc} G_{12}(\varepsilon)  &  F^{(1)}_{12}(\varepsilon) \\ 
F^{(2)}_{12}(\varepsilon) & G^{(h)}_{12}(\varepsilon)\end{array}\right)\nonumber\\
\label{GG4}
\eea
and the analogous expression for ${\hat {\tilde G}}_{12}(\varepsilon)$
with the obvious correspondences between the matrix elements.
The form of Eq. (\ref{GG2}) is convenient for transforming the fermionic propagator to the quasiparticle basis. Indeed, the forward and backward components of the Bogoliubov mean-field quasiparticle propagator ${\tilde G}^{(\pm)}_{\nu\nu'}(\varepsilon)$ defined as
\be
{\tilde G}^{(\eta)}_{\nu\nu'}(\varepsilon) = \frac{\delta_{\nu\nu'}}{\varepsilon - \eta(E_{\nu} - E_0 - i\delta)}
\label{MFG_qp}
\ee
can be obtained by the following transformations:
\bea
{\tilde G}^{(+)}_{\nu\nu'}(\varepsilon) = \sum\limits_{12} \Bigl(U^{\dagger}_{\nu 1} \ \ \  V^{\dagger}_{\nu 1} \Bigr) {\hat {\tilde G}}_{12}(\varepsilon) 
\left( \begin{array}{c} U_{2\nu'} \\ V_{2\nu'} \end{array}\right)\nonumber\\
{\tilde G}^{(-)}_{\nu\nu'}(\varepsilon) = \sum\limits_{12} \Bigl(V^{T}_{\nu 1} \ \ \  U^{T}_{\nu 1} \Bigr) {\hat {\tilde G}}_{12}(\varepsilon) 
\left( \begin{array}{c} V^{\ast}_{2\nu'} \\ U^{\ast}_{2\nu'} \end{array}\right),
\label{GGq}
\eea
that can be verified with the aid of Eqs. (\ref{UV}). The same transformations of the exact propagator of Eq. (\ref{GG1}) result in:
\be
{G}^{(\eta)}_{\nu\nu'}(\varepsilon) = \sum\limits_n\frac{S^{\eta(n)}_{\nu\nu'}}{\varepsilon - \eta(E_{n} - E_0 - i\delta)}
\label{G_qp}
\ee
with the residues $S^{+(n)}_{\nu\nu'} = \langle 0|\alpha_{\nu}|n\rangle\langle n|\alpha^{\dagger}_{\nu'}|0\rangle$ and $S^{-(m)}_{\nu\nu'} = \langle 0|\alpha_{\nu}|m\rangle\langle m|\alpha^{\dagger}_{\nu'}|0\rangle$ formally distinguished by the fact that the states $|n\rangle$ belong to the $(N+1)$-particle system and the states $|m\rangle$ are associated with the $(N-1)$-particle system. This difference will also be neglected in the following.

We can use the advantage of the two-component structure and of the simple form of the fermionic propagator in the quasiparticle basis to generate and solve the equation of motion for this propagator. While the EOM for $G(\varepsilon) = G^{(11)}(\varepsilon)$ is discussed in Subsection \ref{normal}, here we realize that the EOM's for other three components of the propagator ${\hat G}(\varepsilon)$ are needed to complete the system. The $G^{(22)}(\varepsilon) = G^{(h)}(\varepsilon)$ is the hole propagator, and its EOM can be obtained from the one for the $G^{(11)}(\varepsilon)$ by conjugation. The anomalous Green functions $G^{(12)}(\varepsilon) = F^{(1)}(\varepsilon)$ and $G^{(21)}(\varepsilon) = F^{(2)}(\varepsilon)$ require special consideration.

Let us consider first the component $F^{(1)}$
\be
F^{(1)}_{11'}(t-t') = -i\langle T\psi_1(t)\psi_{1'}(t')\rangle .
\label{F1}
\ee
The first EOM for this quantity is again generated by the differentiation with respect to $t$ variable, which leads to:
\be
(i\partial_t - \varepsilon_1)F^{(1)}_{11'}(t-t') = \frac{i}{2}\sum\limits_{ikl}{\bar v}_{i1kl} \langle T(\psi^{\dagger}_i\psi_l\psi_k)(t)\psi_{1'}(t')\rangle,
\ee
while the differentiation with respect to $t'$ yields:
\bea
(i\partial_t - \varepsilon_1)F^{(1)}_{11'}(t-t')(-i\overleftarrow{\partial_{t'}} + \varepsilon_{1'}) = T_{11'}^{(1)}(t-t')   
\label{F1init}
\eea
with the new $T$-matrix
\bea
T_{11'}^{(1)}&(t-t')&  = T_{11'}^{(1)0}(t-t') + T_{11'}^{(1)r}(t-t')\nonumber\\
T_{11'}^{(1)0}&(t-t')& = -\delta(t-t')\langle[[V,\psi_1],\psi_{1'}]_+\rangle\nonumber\\
T_{11'}^{(1)r}&(t-t')& = i\langle T[V,\psi_1](t)[V,\psi_{1'}](t')\rangle ,\nonumber\\
\eea
where the superscript $(1)$ indicates that this $T$-matrix is associated with the anomalous $F^{(1)}$ Green function. The Fourier transformation of Eq. (\ref{F1init}) to the energy domain gives:
\be
F^{(1)}_{11'}(\varepsilon) = \sum\limits_{22'}G^0_{12}(\varepsilon)T_{22'}^{(1)}(\varepsilon)G^{(h)0}_{2'1}(\varepsilon),
\label{F1T}
\ee 
where we defined the free hole propagator
\be
G^{0(h)}_{11'}(\varepsilon) = -G_{1'1}(-\varepsilon) = \frac{\delta_{11'}}{\varepsilon+ \varepsilon_1}, 
\ee 
in addition to the free particle propagator introduced after Eq. (\ref{spEOM3}).
Eq. (\ref{F1T}) defines the general structure of the anomalous propagator $F^{(1)}$: it begins with the free normal particle propagator and ends with the free normal hole one. The $T$-matrix $T^{(1)}$ should, therefore, include all the processes transforming a particle to a hole and a Cooper pair which joins the pairing condensate. 

The static kernel  $T^{(1)0}$ in the energy domain can be calculated straightforwardly:
\bea
T_{11'}^{(1)0} = -\frac{1}{2}\sum\limits_{ikl}{\bar v}_{i1kl}\langle[\psi^{\dagger}_i\psi_l\psi_k,\psi_{1'}]_+\rangle = \nonumber\\
= -\frac{1}{2}\sum\limits_{kl}{\bar v}_{1'1kl}\langle \psi_l\psi_k \rangle = \Delta_{11'}
\label{T10}
\eea
being just the conventional static pairing gap $\Delta$. The dynamical kernel reads:
\bea
T_{11'}^{(1)r}(t-t') = \nonumber\\
= \frac{i}{4}\sum\limits_{ikl}\sum\limits_{mnq}{\bar v}_{i1kl}\langle T(\psi^{\dagger}_i\psi_l\psi_k )(t)(\psi^{\dagger}_m\psi_q\psi_n)(t')\rangle {\bar v}_{m1'nq}.
\label{T1dyn}
\nonumber\\
\eea
This kernel can be then treated in a desirable approximation in full analogy to the case of $T^r$.
Remarkably, the $T$-matrix equation (\ref{F1T}) does not have the free part, which indicates that the anomalous propagator does not exist in free space and represents purely in-medium phenomenon.

The $T$-matrix equations  for the remaining components $F^{(2)}$ and $G^{(h)}$ of the fermionic propagator (\ref{GG4}) can be generated with the same EOM technique, that yields:
\bea
F^{(2)}_{11'}(\varepsilon) &=& \sum\limits_{22'}G^{(h)0}_{12}(\varepsilon)T^{(2)}_{22'}(\varepsilon)G^{0}_{2'1'}(\varepsilon)
\label{F2T}
\\
G^{(h)}_{11'}(\varepsilon) &=& G^{(h)0}_{11'}(\varepsilon) + \sum\limits_{22'}G^{(h)0}_{12}(\varepsilon)T^{(h)}_{22'}(\varepsilon)G^{(h)0}_{2'1'}(\varepsilon).\nonumber\\
\label{GhT}
\eea
It is easy to verify that Eqs. (\ref{spEOM3},\ref{F1T},\ref{F2T},\ref{GhT}) can be combined into one $2\times 2$ matrix equation:
\bea
\left(\begin{array}{cc}G_{11'}(\varepsilon) & F^{(1)}_{11'}(\varepsilon) \\ F^{(2)}_{11'}(\varepsilon) & G^{(h)}_{11'}(\varepsilon)\end{array}\right) = \left(\begin{array}{cc}G^0_{11'}(\varepsilon) & 0 \\ 0 & G^{(h)0}_{11'}(\varepsilon)\end{array}\right) + \nonumber\\
+ \sum\limits_{22'}\left(\begin{array}{cc}G^0_{12}(\varepsilon) & 0 \\ 0 & G^{(h)0}_{12}(\varepsilon)\end{array}\right)\left(\begin{array}{cc}T_{22'}(\varepsilon) & T^{(1)}_{22'}(\varepsilon) \\ T^{(2)}_{22'}(\varepsilon) & T^{(h)}_{22'}(\varepsilon) \end{array}\right)\times\nonumber\\
\times \left(\begin{array}{cc}G^0_{2'1'}(\varepsilon) & 0 \\ 0 & G^{(h)0}_{2'1'}(\varepsilon)\end{array}\right)\nonumber\\
\label{GTgen0}
\eea
or, symbolically,
\be
{\hat G}_{11'}(\varepsilon) = {\hat G}^{0}_{11'}(\varepsilon) + \sum\limits_{22'}{\hat G}^0_{12}(\varepsilon){\hat T}_{22'}(\varepsilon){\hat G}^0_{2'1'}
(\varepsilon)
\label{GTgen}
\ee
with
\bea
{\hat G}^{0}_{11'}(\varepsilon) = \left(\begin{array}{cc}G^0_{11'}(\varepsilon) & 0 \\ 0 & G^{(h)0}_{11'}(\varepsilon)\end{array}\right)\\
{\hat T}_{11'}(\varepsilon) = \left(\begin{array}{cc}T_{11'}(\varepsilon) & T^{(1)}_{11'}(\varepsilon) \\ T^{(2)}_{11'}(\varepsilon) & T^{(h)}_{11'}(\varepsilon) \end{array}\right).
\eea
Now, introducing the irreducible with respect to ${\hat G}^0(\varepsilon)$ self-energy, ${\hat\Sigma}(\varepsilon)$, such as
\be
{\hat T}_{11'}(\varepsilon) = {\hat\Sigma}_{11'}(\varepsilon) + \sum\limits_{22'}{\hat\Sigma}_{12}(\varepsilon){\hat G}^{0}_{22'}(\varepsilon){\hat T}_{2'1'}(\varepsilon),
\ee
or ${\hat\Sigma}_{11'}(\varepsilon) = {\hat T}^{irr}_{11'}(\varepsilon)$ with ${\hat\Sigma}^0_{11'}(\varepsilon) = {\hat T}^{0}_{11'}(\varepsilon)$, 
Eq. (\ref{GTgen}) can be transformed to  the generalized Dyson, or Gor'kov-Dyson, equation:
\be
{\hat G}_{11'}(\varepsilon) = {\hat G}^{0}_{11'}(\varepsilon) + \sum\limits_{22'}{\hat G}^0_{12}(\varepsilon){\hat\Sigma}_{22'}(\varepsilon){\hat G}_{2'1'}(\varepsilon)
\ee
or, explicitly, in the matrix form:
\bea
\left(\begin{array}{cc}G_{11'}(\varepsilon) & F^{(1)}_{11'}(\varepsilon) \\ F^{(2)}_{11'}(\varepsilon) & G^{(h)}_{11'}(\varepsilon)\end{array}\right) = \left(\begin{array}{cc}G^0_{11'}(\varepsilon) & 0 \\ 0 & G^{(h)0}_{11'}(\varepsilon)\end{array}\right) + \nonumber\\
+ \sum\limits_{22'}\left(\begin{array}{cc}G^0_{12}(\varepsilon) & 0 \\ 0 & G^{(h)0}_{12}(\varepsilon)\end{array}\right)\left(\begin{array}{cc}\Sigma_{22'}(\varepsilon) & \Sigma^{(1)}_{22'}(\varepsilon) \\ \Sigma^{(2)}_{22'}(\varepsilon) & \Sigma^{(h)}_{22'}(\varepsilon) \end{array}\right)\times\nonumber\\
\times \left(\begin{array}{cc}G_{2'1'}(\varepsilon) & F^{(1)}_{2'1'}(\varepsilon) \\ F^{(2)}_{2'1'}(\varepsilon) & G^{(h)}_{2'1'}(\varepsilon)\end{array}\right).\nonumber\\
\label{DGE}
\eea
The Gor'kov-Dyson equation in the form (\ref{DGE}) helps reveal the coupling between the normal and anomalous components of the fermionic propagator:
\bea
G = G^0 + G^0\Sigma G + G^0\Sigma^{(1)}F^{(2)}\
\label{G}
\\
F^{(1)} = G^0\Sigma F^{(1)} + G^0\Sigma^{(1)}G^{(h)}\\
F^{(2)} =  G^{(h)0}\Sigma^{(h)} F^{(2)} + G^{(h)0}\Sigma^{(2)}G \\
G^{(h)} =  G^{(h)0} + G^{(h)0}\Sigma^{(h)}G^{(h)} + G^{(h)0}\Sigma^{(2)}F^{(1)} ,
\label{Gh}
\eea
which was not yet obvious in the $T$-matrix equation (\ref{GTgen0}). The system of equations (\ref{G}-\ref{Gh}) is formally similar to that of the Gor'kov theory \cite{Gorkov1958} and the theory of finite Fermi systems \cite{Migdal1967}, where it was obtained for static self-energies (see also Appendix). Eqs. (\ref{G}-\ref{Gh}), thereby, generalize the latter works to the case of the presence of dynamical correlations in the self-energy.

Working in the basis which diagonalized the one-body part of the Hamiltonian, it is convenient to eliminate the free propagators from Eq. (\ref{DGE}) using the decomposition of the self-energy into the static and dynamical terms ${\hat\Sigma}(\varepsilon) = {\hat\Sigma}^0 + {\hat\Sigma}^r(\varepsilon)$. With the mean-field propagator 
defined as the solution of the Gor'kov-Dyson equation with only the static self-energy ${\hat\Sigma}^0$
\be
{\hat {\tilde G}}_{11'}(\varepsilon) = {\hat G}^{0}_{11'}(\varepsilon) + \sum\limits_{22'}{\hat G}^0_{12}(\varepsilon){\hat\Sigma}^0_{22'}{\hat {\tilde G}}_{2'1'}(\varepsilon)
\ee
or
\bea
\left(\begin{array}{cc}{\tilde G}_{11'}(\varepsilon) & {\tilde F}^{(1)}_{11'}(\varepsilon) \\ {\tilde F}^{(2)}_{11'}(\varepsilon) & {\tilde G}^{(h)}_{11'}(\varepsilon)\end{array}\right) = \left(\begin{array}{cc}G^0_{11'}(\varepsilon) & 0 \\ 0 & G^{(h)0}_{11'}(\varepsilon)\end{array}\right) + \nonumber\\
+ \sum\limits_{22'}\left(\begin{array}{cc}G^0_{12}(\varepsilon) & 0 \\ 0 & G^{(h)0}_{12}(\varepsilon)\end{array}\right)\left(\begin{array}{cc}\Sigma^0_{22'}(\varepsilon) & \Sigma^{(1)0}_{22'}(\varepsilon) \\ \Sigma^{(2)0}_{22'}(\varepsilon) & \Sigma^{(h)0}_{22'}(\varepsilon) \end{array}\right)\times\nonumber\\
\times \left(\begin{array}{cc}{\tilde G}_{2'1'}(\varepsilon) & {\tilde F}^{(1)}_{2'1'}(\varepsilon) \\ {\tilde F}^{(2)}_{2'1'}(\varepsilon) & {\tilde G}^{(h)}_{2'1'}(\varepsilon)\end{array}\right)\nonumber\\
\label{DGEMF}
\eea
as the free term, the Gor'kov-Dyson equation for the full quasiparticle propagator takes the form:
\be
{\hat G}_{11'}(\varepsilon) = {\hat {\tilde G}}_{11'}(\varepsilon) + \sum\limits_{22'}{\hat {\tilde G}}_{12}(\varepsilon){\hat\Sigma}^r_{22'}(\varepsilon){\hat G}_{2'1'}(\varepsilon).
\ee
The component structure of this equation is, explicitly,
\bea
\left(\begin{array}{cc}G_{11'}(\varepsilon) & F^{(1)}_{11'}(\varepsilon) \\ F^{(2)}_{11'}(\varepsilon) & G^{(h)}_{11'}(\varepsilon)\end{array}\right) = \left(\begin{array}{cc}{\tilde G}_{11'}(\varepsilon) & {\tilde F}^{(1)}_{11'}(\varepsilon) \\ {\tilde F}^{(2)}_{11'}(\varepsilon) & {\tilde G}^{(h)}_{11'}(\varepsilon)\end{array}\right) + \nonumber\\
+ \sum\limits_{22'}\left(\begin{array}{cc}{\tilde G}_{12}(\varepsilon) & {\tilde F}^{(1)}_{12}(\varepsilon) \\ {\tilde F}^{(2)}_{12}(\varepsilon) & {\tilde G}^{(h)}_{12}(\varepsilon)\end{array}\right) \left(\begin{array}{cc}\Sigma^r_{22'}(\varepsilon) & \Sigma^{(1)r}_{22'}(\varepsilon) \\ \Sigma^{(2)r}_{22'}(\varepsilon) & \Sigma^{(h)r}_{22'}(\varepsilon) \end{array}\right)\times\nonumber\\
\times \left(\begin{array}{cc}G_{2'1'}(\varepsilon) & F^{(1)}_{2'1'}(\varepsilon) \\ F^{(2)}_{2'1'}(\varepsilon) & G^{(h)}_{2'1'}(\varepsilon)\end{array}\right),\nonumber\\
\label{DGE1}
\eea
where the non-diagonal structure of the free mean-field term induces couplings to all types of the energy-dependent self-energies for each propagator component.
Since the full and the mean-field propagators, respectively, 
\bea
{\hat {G}}_{11'}(\varepsilon) \equiv
\left(\begin{array}{cc}{G}_{11'}(\varepsilon) & {F}^{(1)}_{11'}(\varepsilon) \\ {F}^{(2)}_{11'}(\varepsilon) & {G}^{(h)}_{11'}(\varepsilon)\end{array}\right)\\
{\hat {\tilde G}}_{11'}(\varepsilon) \equiv
\left(\begin{array}{cc}{\tilde G}_{11'}(\varepsilon) & {\tilde F}^{(1)}_{11'}(\varepsilon) \\ {\tilde F}^{(2)}_{11'}(\varepsilon) & {\tilde G}^{(h)}_{11'}(\varepsilon)\end{array}\right),
\eea
are defined by the spectral representations of Eqs. (\ref{GG1},\ref{GG2}), the Gor'kov-Dyson equation (\ref{DGE1}) can be transformed to the quasiparticle basis applying the transformations introduced in Eqs. (\ref{GGq}) to Eqs. (\ref{GG1},\ref{GG2}). These operations yield the Gor'kov-Dyson equation in the quasiparticle basis:
\be
G^{(\eta)}_{\nu\nu'}(\varepsilon) = {\tilde G}^{(\eta)}_{\nu\nu'}(\varepsilon) + \sum\limits_{\mu\mu'}{\tilde G}^{(\eta)}_{\nu\mu}(\varepsilon)\Sigma^{r(\eta)}_{\mu\mu'}(\varepsilon)G^{(\eta)}_{\mu'\nu'}(\varepsilon), 
\label{Dyson_qp}
\ee
with $\eta = +$ and $\eta = -$, the quasiparticle forward and backward components isolated by the first and the second transformations of Eq. (\ref{GGq}), respectively. The components of the 
dynamical kernel are, accordingly, transformed to the quasiparticle space as
\bea
\Sigma^{r(+)}_{\mu\mu'}(\varepsilon) &=& \sum\limits_{12} \Bigl(U^{\dagger}_{\mu 1} \ \ \  V^{\dagger}_{\mu 1} \Bigr) \left( \begin{array}{cc} \Sigma^r_{12}(\varepsilon) &  \Sigma^{(1)r}_{12}(\varepsilon)\\ \Sigma^{(2)r}_{12}(\varepsilon) & \Sigma^{(h)r}_{12}(\varepsilon)\end{array}\right)
\left( \begin{array}{c} U_{2\mu'} \\ V_{2\mu'} \end{array}\right)  \nonumber\\
&=& \sum\limits_{12} \Bigl( U^{\dagger}_{\mu 1}\Sigma^r_{12}U_{2\mu'} + U^{\dagger}_{\mu 1}\Sigma^{(1)r}_{12}V_{2\mu'} \nonumber\\
&+& V^{\dagger}_{\mu 1}\Sigma^{(2)r}_{12}U_{2\mu'} + V^{\dagger}_{\mu 1}\Sigma^{(h)r}_{12}V_{2\mu'} \Bigr),
\label{Sigma+}
\eea
\bea
\Sigma^{r(-)}_{\mu\mu'}(\varepsilon) &=& \sum\limits_{12} \Bigl(V^{T}_{\mu 1} \ \ \  U^{T}_{\mu 1} \Bigr) \left( \begin{array}{cc} \Sigma^r_{12}(\varepsilon) &  \Sigma^{(1)r}_{12}(\varepsilon)\\ \Sigma^{(2)r}_{12}(\varepsilon) & \Sigma^{(h)r}_{12}(\varepsilon)\end{array}\right)
\left( \begin{array}{c} V^{\ast}_{2\mu'} \\ U^{\ast}_{2\mu'} \end{array}\right)  \nonumber\\
&=& \sum\limits_{12} \Bigl( V^{T}_{\mu 1}\Sigma^r_{12}V^{\ast}_{2\mu'} + V^{T}_{\mu 1}\Sigma^{(1)r}_{12}U^{\ast}_{2\mu'} \nonumber\\
&+& U^{T}_{\mu 1}\Sigma^{(2)r}_{12}V^{\ast}_{2\mu'} + U^{T}_{\mu 1}\Sigma^{(h)r}_{12}U^{\ast}_{2\mu'} \Bigr).
\label{Sigma-}
\eea
Eqs. (\ref{Dyson_qp} - \ref{Sigma-}), together with the mean-field propagator of Eq. (\ref{MFG_qp}), thereby, completely define the quasiparticle propagator in a superfluid fermionic system. Notice here that in the implementations of the Gor'kov-Dyson equation it is convenient to use the spectral form of Eq. (\ref{G_qp}) for the full propagator, that reduces the problem to finding its poles and the corresponding residues, or the spectroscopic factors. The considerable advantage of transforming the  Gor'kov-Dyson equation to the quasiparticle basis is that in this basis one deals with, formally, only two components of the propagator, instead of four of them in the single-particle basis. Moreover, with the relaxed particle number conservation condition, which is a feature of the HFB approach and a good approximation of large-$N$ fermionic systems, one can notice that in Eq. (\ref{G_qp}) $S^{+(n)}_{\nu\nu'} = S^{-(n)}_{\nu\nu'}$, so that the solutions for $\eta = +$ and $\eta = -$ are doubling each other also in the theory extended by dynamical correlations in the self-energy. This means that only one of the Eqs. (\ref{Dyson_qp}) needs to be solved, that further reduces the computation effort by a factor of two.

\subsection{The self-energy in the intermediate and strong coupling regimes: superfluid PVC, or quasiparticle-vibration coupling (QVC)}
\label{SE_qp}

While the static part of the self-energy is determined unambiguously by Eqs. (\ref{MF},\ref{T10}), it still depends explicitly on the one-fermion normal and pairing densities, which are determined by the static limit of the full quasiparticle propagator $\hat G$, according to Eq. (\ref{Valatin}). Thereby, in a self-consistent theory the static self-energy depends on the approximation made for the quasiparticle propagator, i.e., on the approximation for the dynamical, or energy-dependent, self-energy ${\hat\Sigma}^{r}_{12}(\varepsilon)$. Below we discuss in detail this part of the self-energy bearing in mind its matrix structure introduced above:
\be
{\hat\Sigma}^{r}_{12}(\varepsilon) = \left( \begin{array}{cc} \Sigma^r_{12}(\varepsilon) &  \Sigma^{(1)r}_{12}(\varepsilon)\\ \Sigma^{(2)r}_{12}(\varepsilon) & \Sigma^{(h)r}_{12}(\varepsilon)\end{array}\right).
\ee
The PVC approach was derived for the $\Sigma^r_{12}(\varepsilon)$ component of ${\hat\Sigma}^{r}_{12}(\varepsilon)$ in Section \ref{normal} with the aid of a cluster decomposition of the three-fermion propagator.  However, in modeling the superfluid phase, where the particle number conservation is relaxed, the component $\Sigma^r_{12}(\varepsilon)$ should be extended for the inclusion of the propagators via the states with varied particle number. Keeping all terms containing up to two-fermion correlation functions and neglecting the uncorrelated term, which is supposed to be of little significance in the intermediate and strong coupling regimes, one obtains for the irreducible part of the dynamical $T$-matrix $T^{r}$:
\bea
\Sigma_{11'}^{r}(t-t') = T_{11'}^{r;irr}(t-t') = \nonumber\\
= -\frac{i}{4}\sum\limits_{ikl}\sum\limits_{pqr}{\bar v}_{1ikl}\langle T(\psi^{\dagger}_i\psi_l\psi_k )(t)(\psi^{\dagger}_q\psi^{\dagger}_p\psi_r)(t')\rangle^{irr} {\bar v}_{pqr1'} \approx \nonumber\\
\approx -\frac{i}{4}\sum\limits_{ikl}\sum\limits_{pqr}{\bar v}_{1ikl}
\Bigl[
\langle T\psi_k(t)\psi^{\dagger}_q(t')\rangle\langle T(\psi^{\dagger}_i\psi_l)(t)(\psi^{\dagger}_p\psi_r)(t')\rangle - \nonumber\\
- \langle T\psi_l(t)\psi^{\dagger}_q(t')\rangle\langle T(\psi^{\dagger}_i\psi_k)(t)(\psi^{\dagger}_p\psi_r)(t')\rangle + \nonumber\\
+ \langle T\psi_l(t)\psi^{\dagger}_p(t')\rangle\langle T(\psi^{\dagger}_i\psi_k)(t)(\psi^{\dagger}_q\psi_r)(t')\rangle - \nonumber\\
- \langle T\psi_k(t)\psi^{\dagger}_p(t')\rangle\langle T(\psi^{\dagger}_i\psi_l)(t)(\psi^{\dagger}_q\psi_r)(t')\rangle + \nonumber\\
+ \langle T\psi^{\dagger}_i(t)\psi_r(t')\rangle\langle T(\psi_l\psi_k)(t)(\psi^{\dagger}_q\psi^{\dagger}_p)(t')\rangle + \nonumber\\
+ \langle T\psi_k(t)\psi_r(t')\rangle\langle T(\psi^{\dagger}_i\psi_l)(t)(\psi^{\dagger}_q\psi^{\dagger}_p)(t')\rangle - \nonumber\\
- \langle T\psi_l(t)\psi_r(t')\rangle\langle T(\psi^{\dagger}_i\psi_k)(t)(\psi^{\dagger}_q\psi^{\dagger}_p)(t')\rangle + \nonumber\\
+ \langle T\psi^{\dagger}_i(t)\psi^{\dagger}_q(t')\rangle\langle T(\psi_l\psi_k)(t)(\psi^{\dagger}_p\psi_r)(t')\rangle - \nonumber\\
- \langle T\psi^{\dagger}_i(t)\psi^{\dagger}_p(t')\rangle\langle T(\psi_l\psi_k)(t)(\psi^{\dagger}_q\psi_r)(t')\rangle
\Bigr]
{\bar v}_{pqr1'}.
\label{Tdyn_clus}
\nonumber\\
\eea
The first four terms in the square brackets form the fully antisymmetrized product of the single-fermion normal propagator and the particle-phonon response, and the fifth term is the single-hole normal propagator coupled to the two-fermion Green function. These terms are the same as we had in Section \ref{normal} for the particle number conserving normal phase. As we relax the condition of the particle number conservation here, the additional four terms with anomalous one-fermion and two-fermion propagators appear in $\Sigma^r$.
Employing the definitions of $F^{(1)}$ and $R$ of Eqs. (\ref{F1},\ref{phresp}), and adding the second anomalous fermionic propagator $F^{(2)}$ 
\be
F^{(2)}_{11'}(t-t') = -i\langle T\psi^{\dagger}_1(t)\psi^{\dagger}_{1'}(t')\rangle
\label{F2}
\ee
as well as the double-anomalous fermionic pair propagators
\bea
{G}^{(01)}_{12,1'2'}(t-t') =
-i\langle T(\psi^{\dagger}_1\psi_2)(t)(\psi_{2'}\psi_{1'})(t')\rangle,
\label{G01abnor}
\\
{G}^{(10)}_{12,1'2'}(t-t') =
-i\langle T(\psi_1\psi_2)(t)(\psi^{\dagger}_{2'}\psi_{1'})(t')\rangle,
\label{G10abnor}
\\
{G}^{(11)}_{12,1'2'}(t-t') =
-i\langle T(\psi_1\psi_2)(t)(\psi_{2'}\psi_{1'})(t')\rangle,
\label{G11abnor}
\\
{G}^{(02)}_{12,1'2'}(t-t') =
-i\langle T(\psi^{\dagger}_1\psi_2)(t)(\psi^{\dagger}_{2'}\psi^{\dagger}_{1'})(t')\rangle,
\label{G02abnor}
\\
{G}^{(20)}_{12,1'2'}(t-t') =
-i\langle T(\psi^{\dagger}_1\psi^{\dagger}_2)(t)(\psi^{\dagger}_{2'}\psi_{1'})(t')\rangle,
\label{G20abnor}
\\
{G}^{(22)}_{12,1'2'}(t-t') =
-i\langle T(\psi^{\dagger}_1\psi^{\dagger}_2)(t)(\psi^{\dagger}_{2'}\psi^{\dagger}_{1'})(t')\rangle,
\label{G22abnor}
\eea
one can recast the dynamical self-energy $\Sigma_{11'}^{r}(t-t') $ in the following form: 
\bea
{\Sigma}_{11'}^{r}(t-t') = \frac{i}{4}\sum\limits_{ikl}\sum\limits_{prq}{\bar v}_{1ikl}\Bigl[ 
4G_{kq}(t-t')R_{il,rp}(t-t') + \nonumber\\
+ G^{(h)}_{ir}(t-t'){G}_{lk,pq}(t-t') + 2F^{(1)}_{kr}(t-t')G^{(02)}_{il,pq}(t-t') + \nonumber\\
+ 2F^{(2)}_{iq}(t-t')G^{(10)}_{lk,rp}(t-t')
\Bigr] {\bar v}_{pqr1'}.\nonumber\\
\label{Sigma_qp}
\eea
Notice that here and below we use the redefined two-fermion propagator $iG_{12,1'2'}(t-t') \to G_{12,1'2'}(t-t')$, i.e., introduce an additional factor $i$ in the right hand side of Eq. (\ref{ppGF}). This allows one to treat both the particle-hole and the particle-particle propagator in a unified way. With this redefinition and with $G^{(h)}(1,2) = -G(2,1)$ it is easy to verify that the first two terms in Eq. (\ref{Sigma_qp}) are the same as those in Eq. (\ref{SEirr2}). Otherwise, compared to the latter expression, in Eq.  (\ref{Sigma_qp}) we dropped the uncorrelated terms, which are relatively unimportant in intermediate and strong coupling regimes, and added the two other terms with the anomalous propagators arising from the decomposition of Eq. (\ref{Tdyn_clus}).
To transform ${\Sigma}_{11'}^{r}(t-t')$ of Eq. (\ref{Sigma_qp}), with the new last two terms, to the energy domain, the following spectral expansions of these propagators are helpful:
\bea
F^{(1)}_{11'}(t-t') &=& -i\theta(t-t')\sum\limits_n e^{-i\epsilon_n(t-t')}\eta^n_1\chi^n_{1'} +\nonumber\\
&+& i\theta(t'-t)\sum\limits_m e^{i\epsilon_m(t-t')}\chi^m_1\eta^m_{1'} 
\label{F1spec}
\eea
\bea
F^{(2)}_{11'}(t-t') &=& -i\theta(t-t')\sum\limits_n e^{-i\epsilon_n(t-t')}\chi^{n\ast}_1\eta^{n\ast}_{1'} +\nonumber\\
&+& i\theta(t'-t)\sum\limits_m e^{i\epsilon_m(t-t')}\eta^{m\ast}_1\chi^{m\ast}_{1'} 
\label{F2spec}
\eea
\bea
{G}^{(10)}_{12,1'2'}(t-t') &=& -i\theta(t-t')\sum\limits_{\mu} e^{-i\omega_{\mu}(t-t')}\alpha^{\mu}_{21}\rho^{\mu\ast}_{2'1'} -\nonumber\\
&-& i\theta(t'-t)\sum\limits_{\varkappa} e^{i\omega_{\varkappa}(t-t')}\beta^{\varkappa\ast}_{12}\rho^{\varkappa}_{1'2'},  
\label{G10abnor_spec}
\eea
\bea
{G}^{(02)}_{12,1'2'}(t-t') &=& -i\theta(t-t')\sum\limits_{\mu} e^{-i\omega_{\mu}(t-t')}\rho^{\mu}_{21}\alpha^{\mu\ast}_{2'1'} -\nonumber\\
&-& i\theta(t'-t)\sum\limits_{\varkappa} e^{i\omega_{\varkappa}(t-t')}\rho^{\varkappa\ast}_{12}\beta^{\varkappa}_{1'2'},  
\label{G02abnor_spec}
\eea
where the matrix elements $\chi, \eta, \alpha, \beta$ were defined in Eqs. (\ref{etachi},\ref{alphabeta}) and $\rho$ was introduced in Eq. (\ref{rho}). Here, as mentioned above, to have non-vanishing anomalous propagators, we relax the particle number conservation condition on both the ground and excited states, so that the particle number is conserved only on average. In Eqs. (\ref{F1spec} -- \ref{G02abnor_spec}) we have introduced the single-particle energy differences: $\epsilon_n = E_n - E_0$ and  
the transition frequencies $\omega_{\mu} = E_{\mu} - E_0$.  We still keep different indices for numbering the intermediate states in odd and even nuclei in the positive-frequency and negative-frequency components of the propagators, such as the pairs $n-m$ and $\mu-\varkappa$, but keep in mind that the states with variable particle numbers are treated on equal grounds.

With the help of Eqs. (\ref{F1spec} -- \ref{G02abnor_spec}), completed by the similar expressions for the particle and hole propagators as well as for the particle-hole and the particle-particle propagators, 
\bea
G_{11'}(t-t') &=& -i\theta(t-t')\sum\limits_n e^{-i\epsilon_n(t-t')}\eta^{n}_1\eta^{n\ast}_{1'} +\nonumber\\
&+& i\theta(t'-t)\sum\limits_m e^{i\epsilon_m(t-t')}\chi^{m}_1\chi^{m\ast}_{1'} 
\label{Gspec}
\eea
\bea
G^{(h)}_{11'}(t-t') &=& -i\theta(t-t')\sum\limits_n e^{-i\epsilon_n(t-t')}\chi^{n\ast}_1\chi^n_{1'} +\nonumber\\
&+& i\theta(t'-t)\sum\limits_m e^{i\epsilon_m(t-t')}\eta^{m\ast}_1\eta^m_{1'} 
\label{Ghspec}
\eea
\bea
{R}_{12,1'2'}(t-t') &=& -i\theta(t-t')\sum\limits_{\nu} e^{-i\omega_{\nu}(t-t')}\rho^{\nu}_{21}\rho^{\nu\ast}_{2'1'} -\nonumber\\
&-& i\theta(t'-t)\sum\limits_{\nu} e^{i\omega_{\nu}(t-t')}\rho^{\nu\ast}_{12}\rho^{\nu}_{1'2'}  
\label{R_spec}
\eea
\bea
G_{12,1'2'}(t-t') &=& -i\theta(t-t')\sum\limits_{\mu} e^{-i\omega_{\mu}(t-t')}\alpha^{\mu}_{21}\alpha^{\mu\ast}_{2'1'} -\nonumber\\
&-& i\theta(t'-t)\sum\limits_{\varkappa} e^{i\omega_{\varkappa}(t-t')}\beta^{\varkappa\ast}_{12}\beta^{\varkappa}_{1'2'},  
\label{G2_spec}
\eea

the Fourier image of ${\Sigma}^{r}$ takes the form:
\bea
\Sigma^r_{11'}(\varepsilon) = \int\limits_{-\infty}^{\infty}d\tau e^{i\varepsilon\tau}{\Sigma}_{11'}^{r}(\tau) = \nonumber \\
= \sum\limits_{33'} \Bigl[ 
\sum\limits_{\nu n}\frac{{g}_{13}^{\nu}\eta_3^{n}\eta_{3'}^{n\ast}{g}_{1'3'}^{\nu\ast}}{\varepsilon  - \epsilon_n - \omega_{\nu} + i\delta} +
\sum\limits_{\nu m} \frac{ g_{31}^{\nu\ast}\chi_3^{m}\chi_{3'}^{m\ast}g_{3'1'}^{\nu}}{\varepsilon + \epsilon_m + \omega_{\nu} - i\delta} 
+ \nonumber\\
+ \sum\limits_{\mu m} \frac{ \gamma_{13}^{\mu(+)}\chi_3^{m\ast}\chi_{3'}^{m}\gamma_{1'3'}^{\mu(+)\ast}}{\varepsilon  - \epsilon_m - \omega_{\mu} + i\delta} 
+ \sum\limits_{\varkappa n}\frac{{\gamma}_{31}^{\varkappa(-)\ast}\eta_3^{n\ast}\eta_{3'}^n{\gamma}_{3'1'}^{\varkappa(-)}}{\epsilon  + \epsilon_n + \omega_{\varkappa} - i\delta} \nonumber\\
+
\sum\limits_{\mu n}\frac{{g}_{13}^{\mu}\eta_3^{n}\chi_{3'}^{n}{\gamma}_{1'3'}^{\mu
(+)\ast}}{\varepsilon  - \epsilon_n - \omega_{\mu} + i\delta} +
\sum\limits_{\varkappa m} \frac{ g_{31}^{\varkappa\ast}\chi_3^{m}\eta_{3'}^{m}\gamma_{3'1'}^{\varkappa
(-)}}{\varepsilon + \epsilon_m + \omega_{\varkappa} - i\delta} 
+ \nonumber\\
+ \sum\limits_{\mu n} \frac{ \gamma_{13}^{\mu(+)}\chi_3^{n\ast}\eta_{3'}^{n\ast}g_{1'3'}^{\mu\ast}}{\varepsilon  - \epsilon_n - \omega_{\mu} + i\delta} 
+ \sum\limits_{\varkappa m}\frac{{\gamma}_{31}^{\varkappa(-)\ast}\eta_3^{m\ast}\chi_{3'}^
{m\ast}{g}_{3'1'}^{\varkappa}}{\varepsilon  + \epsilon_m + \omega_{\varkappa} - i\delta} 
\Bigr],\nonumber\\
\label{Sigma_r}
\eea
where the definition of the normal and pairing phonon vertices (\ref{vert_pp},\ref{vert_ph}) were employed.

Derivations analogous to that for $\Sigma^r$ can be conducted for the other components of the dynamical self-energy. To get an expression for  
$\Sigma^{(1)r}$, one can start with the exact $T$-matrix $T^{(1)r}$ of Eq. (\ref{T1dyn}) and approximate its irreducible part by the cluster decomposition retaining all terms up to those with two-fermion correlation functions. Again, neglecting the uncorrelated terms, one gets:
\bea
 \Sigma_{11'}^{(1)r}(t-t') = T_{11'}^{(1)r;irr}(t-t') =\nonumber\\
= \frac{i}{4}\sum\limits_{ikl}\sum\limits_{mnq}{\bar v}_{i1kl}\langle T(\psi^{\dagger}_i\psi_l\psi_k )(t)(\psi^{\dagger}_m\psi_q\psi_n)(t')\rangle^{irr} {\bar v}_{m1'nq} \approx \nonumber\\
\approx \frac{i}{4}\sum\limits_{ikl}\sum\limits_{mnq}{\bar v}_{i1kl}
\Bigl[
\langle T\psi_k(t)\psi_n(t')\rangle\langle T(\psi^{\dagger}_i\psi_l)(t)(\psi^{\dagger}_m\psi_q)(t')\rangle + \nonumber\\
+ \langle T\psi_l(t)\psi_q(t')\rangle\langle T(\psi^{\dagger}_i\psi_k)(t)(\psi^{\dagger}_m\psi_n)(t')\rangle - \nonumber\\
- \langle T\psi_k(t)\psi_q(t')\rangle\langle T(\psi^{\dagger}_i\psi_l)(t)(\psi^{\dagger}_m\psi_n)(t')\rangle - \nonumber\\
- \langle T\psi_l(t)\psi_n(t')\rangle\langle T(\psi^{\dagger}_i\psi_k)(t)(\psi^{\dagger}_m\psi_q)(t')\rangle + \nonumber\\
+ \langle T\psi^{\dagger}_i(t)\psi^{\dagger}_m(t')\rangle\langle T(\psi_l\psi_k)(t)(\psi_q\psi_n)(t')\rangle - \nonumber\\
- \langle T\psi_l(t)\psi^{\dagger}_m(t')\rangle\langle T(\psi^{\dagger}_i\psi_k)(t)(\psi_q\psi_n)(t')\rangle + \nonumber\\
+ \langle T\psi_k(t)\psi^{\dagger}_m(t')\rangle\langle T(\psi^{\dagger}_i\psi_l)(t)(\psi_q\psi_n)(t')\rangle - \nonumber\\
- \langle T\psi^{\dagger}_i(t)\psi_q(t')\rangle\langle T(\psi_l\psi_k)(t)(\psi^{\dagger}_m\psi_n)(t')\rangle + \nonumber\\
+ \langle T\psi^{\dagger}_i(t)\psi_n(t')\rangle\langle T(\psi_l\psi_k)(t)(\psi^{\dagger}_m\psi_q)(t')\rangle
\Bigr]
{\bar v}_{m1'nq}.
\label{T1dyn_clus}
\nonumber\\
\eea
With the definitions introduced above, the $\Sigma^{(1)r}$ component of the dynamical self-energy can be rewritten as
\bea
{\Sigma}_{11'}^{(1)r}(t-t') = -\frac{i}{4}\sum\limits_{ikl}\sum\limits_{prq}{\bar v}_{i1kl}\Bigl[ 
4F^{(1)}_{kr}(t-t')R_{il,qp}(t-t') + \nonumber\\
+ F^{(2)}_{ip}(t-t'){G}^{(11)}_{lk,rq}(t-t') + 2G_{kp}(t-t')G^{(01)}_{il,rq}(t-t') + \nonumber\\
+ 2G^{(h)}_{ir}(t-t')G^{(10)}_{lk,qp}(t-t')
\Bigr] {\bar v}_{p1'rq}.\nonumber\\
\eea
The transformation of ${\Sigma}_{11'}^{(1)r}(t-t')$ to the energy domain requires the additional spectral expansions of the anomalous propagators:
\bea
{G}^{(01)}_{12,1'2'}(t-t') &=& -i\theta(t-t')\sum\limits_{\mu} e^{-i\omega_{\mu}(t-t')}\rho^{\mu}_{21}\beta^{\mu\ast}_{2'1'} -\nonumber\\
&-& i\theta(t'-t)\sum\limits_{\varkappa} e^{i\omega_{\varkappa}(t-t')}\rho^{\varkappa\ast}_{12}\alpha^{\varkappa}_{1'2'}  
\label{G01abnor_spec}
\eea
\bea
{G}^{(11)}_{12,1'2'}(t-t') &=& -i\theta(t-t')\sum\limits_{\mu} e^{-i\omega_{\mu}(t-t')}\alpha^{\mu}_{21}\beta^{\mu\ast}_{2'1'} -\nonumber\\
&-& i\theta(t'-t)\sum\limits_{\varkappa} e^{i\omega_{\varkappa}(t-t')}\beta^{\varkappa\ast}_{12}\alpha^{\varkappa}_{1'2'},  
\label{G11abnor_spec}
\eea
so that
the Fourier image of $\Sigma^{(1)r}$ then reads:
\bea
\Sigma^{(1)r}_{11'}(\varepsilon) = \int\limits_{-\infty}^{\infty}d\tau e^{i\varepsilon\tau}{\Sigma}_{11'}^{(1)r}(\tau) =
\nonumber \\
= -\sum\limits_{33'} \Bigl[ 
\sum\limits_{\nu n}\frac{{g}_{13}^{\nu}\eta_3^{n}\chi_{3'}^{n}{g}_{3'1'}^{\nu\ast}}{\varepsilon  - \epsilon_n - \omega_{\nu} + i\delta} +
\sum\limits_{\nu m} \frac{ g_{31}^{\nu\ast}\chi_3^{m}\eta_{3'}^{m}g_{1'3'}^{\nu}}{\varepsilon + \epsilon_m + \omega_{\nu} - i\delta} 
+ \nonumber\\
+ \sum\limits_{\mu n} \frac{ \gamma_{13}^{\mu(+)}\chi_3^{n\ast}\eta_{3'}^{n\ast}\gamma_{3'1'}^{\mu(-)\ast}}{\varepsilon - \epsilon_n - \omega_{\mu} + i\delta} 
+ \sum\limits_{\varkappa m}\frac{{\gamma}_{31}^{\varkappa(-)\ast}\eta_3^{m\ast}\chi_{3'}^{m\ast}{\gamma}_{1'3'}^{\varkappa(+)}}{\varepsilon +  \epsilon_m + \omega_{\varkappa} - i\delta} 
+\nonumber\\
+ \sum\limits_{\mu n} \frac{g_{13}^{\mu}\eta_3^{n}\eta^{n\ast}_{3'}\gamma_{3'1'}^{\mu(-)\ast}}{\varepsilon - \epsilon_n - \omega_{\mu} + i\delta} 
+ \sum\limits_{\varkappa m}\frac{g_{31}^{\varkappa\ast}\chi_3^{m}\chi_{3'}^{m\ast}{\gamma}_{1'3'}^{\varkappa(+)}}{\varepsilon +  \epsilon_m + \omega_{\varkappa} - i\delta} 
+\nonumber\\
+ \sum\limits_{\mu n} \frac{\gamma_{13}^{\mu(+)}\chi_3^{n\ast}\chi^{n}_{3'}g_{3'1'}^{\mu\ast}}{\varepsilon - \epsilon_n - \omega_{\mu} + i\delta} 
+ \sum\limits_{\varkappa m}\frac{\gamma^{\varkappa(-)\ast}_{31}\eta_3^{m\ast}\eta_{3'}^{m}g_{1'3'}^{\varkappa}}{\varepsilon +  \epsilon_m + \omega_{\varkappa} - i\delta} 
\Bigr].\nonumber\\
\label{Sigma_1r}
\eea
\begin{figure*}
\begin{center}
\includegraphics*[scale=0.75]{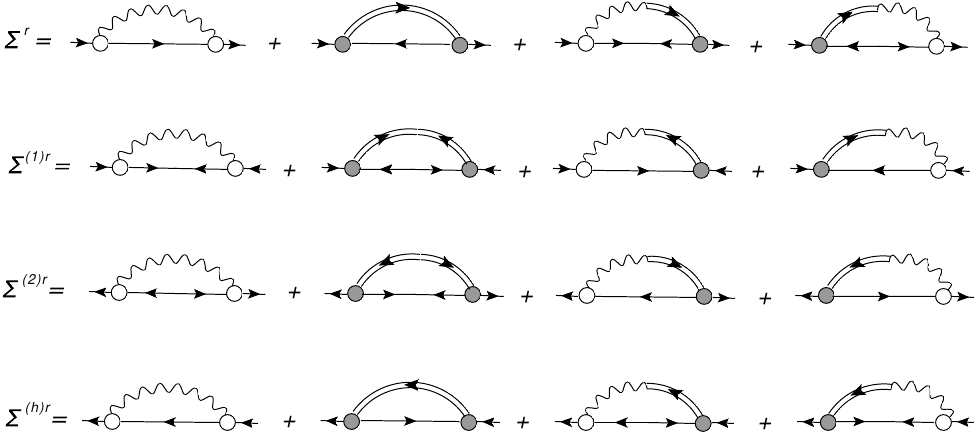}
\end{center}
\caption{The component structure of the QVC self-energy given in Eqs. (\ref{Sigma_r},\ref{Sigma_1r},\ref{Sigma_2r},\ref{Sigma_hr}) in diagrammatic form.}
\label{SEgen}%
\end{figure*}

The remaining components $\Sigma^{(2)}$ and $\Sigma^{(h)}$ are found analogously and read:
\bea
\Sigma^{(2)r}_{11'}(\varepsilon) = \int\limits_{-\infty}^{\infty}d\tau e^{i\varepsilon\tau}{\Sigma}_{11'}^{(2)r}(\tau) =
\nonumber \\
= -\sum\limits_{33'} \Bigl[ 
\sum\limits_{\nu n}\frac{{g}_{31}^{\nu}\chi_3^{n\ast}\eta_{3'}^{n\ast}{g}_{1'3'}^{\nu\ast}}{\varepsilon  - \epsilon_n - \omega_{\nu} + i\delta} +
\sum\limits_{\nu m} \frac{ g_{13}^{\nu\ast}\eta_3^{m\ast}\chi_{3'}^{m\ast}g_{3'1'}^{\nu}}{\varepsilon + \epsilon_m + \omega_{\nu} - i\delta} 
+ \nonumber\\
+ \sum\limits_{\mu n} \frac{ \gamma_{31}^{\mu(-)}\eta_3^{n}\chi_{3'}^{n}\gamma_{1'3'}^{\mu(+)\ast}}{\varepsilon - \epsilon_n - \omega_{\mu} + i\delta} 
+ \sum\limits_{\varkappa m}\frac{{\gamma}_{13}^{\varkappa(+)\ast}\chi_3^{m}\eta_{3'}^{m}{\gamma}_{3'1'}^{\varkappa(-)}}{\varepsilon +  \epsilon_m + \omega_{\varkappa} - i\delta} 
+\nonumber\\
+ \sum\limits_{\mu n} \frac{g_{31}^{\mu}\chi_3^{n\ast}\chi^{n}_{3'}\gamma_{1'3'}^{\mu(+)\ast}}{\varepsilon - \epsilon_n - \omega_{\mu} + i\delta} 
+ \sum\limits_{\varkappa m}\frac{g_{13}^{\varkappa\ast}\eta_3^{m\ast}\eta_{3'}^{m}{\gamma}_{3'1'}^{\varkappa(-)}}{\varepsilon +  \epsilon_m + \omega_{\varkappa} - i\delta} 
+\nonumber\\
+ \sum\limits_{\mu n} \frac{\gamma_{31}^{\mu(-)}\eta_3^{n}\eta^{n\ast}_{3'}g_{1'3'}^{\mu\ast}}{\varepsilon - \epsilon_n - \omega_{\mu} + i\delta} 
+ \sum\limits_{\varkappa m}\frac{\gamma^{\varkappa(+)\ast}_{13}\chi_3^{m}\chi_{3'}^{m\ast}g_{3'1'}^{\varkappa}}{\varepsilon +  \epsilon_m + \omega_{\varkappa} - i\delta} 
\Bigr].\nonumber\\
\label{Sigma_2r}
\eea
\bea
\Sigma^{(h)r}_{11'}(\varepsilon) = \int\limits_{-\infty}^{\infty}d\tau e^{i\varepsilon\tau}{\Sigma}_{11'}^{(h)r}(\tau) = \nonumber \\
= \sum\limits_{33'} \Bigl[ 
\sum\limits_{\nu m}\frac{{g}_{31}^{\nu}\chi_3^{m\ast}\chi_{3'}^{m}{g}_{3'1'}^{\nu\ast}}{\varepsilon  - \epsilon_m - \omega_{\nu} + i\delta} +
\sum\limits_{\nu n} \frac{ g_{13}^{\nu\ast}\eta_3^{n\ast}\eta_{3'}^{n}g_{1'3'}^{\nu}}{\varepsilon + \epsilon_n + \omega_{\nu} - i\delta} 
+ \nonumber\\
+ \sum\limits_{\mu n} \frac{ \gamma_{31}^{\mu(-)}\eta_3^{n}\eta_{3'}^{n\ast}\gamma_{3'1'}^{\mu(-)\ast}}{\varepsilon  - \epsilon_n - \omega_{\mu} + i\delta} 
+ \sum\limits_{\varkappa m}\frac{{\gamma}_{13}^{\varkappa(+)\ast}\chi_3^{m}\chi_{3'}^{m\ast}{\gamma}_{1'3'}^{\varkappa(+)}}{\epsilon  + \epsilon_m + \omega_{\varkappa} - i\delta} \nonumber\\
+
\sum\limits_{\mu n}\frac{g_{31}^{\mu}\chi_3^{n\ast}\eta_{3'}^{n\ast}{\gamma}_{3'1'}^{\mu(-)\ast}}{\varepsilon  - \epsilon_n - \omega_{\mu} + i\delta} +
\sum\limits_{\varkappa m} \frac{ g_{13}^{\varkappa\ast}\eta_3^{m\ast}\chi_{3'}^{m\ast}\gamma_{1'3'}^{\varkappa
(+)}}{\varepsilon + \epsilon_m + \omega_{\varkappa} - i\delta} 
+ \nonumber\\
+ \sum\limits_{\mu n} \frac{ \gamma_{31}^{\mu(-)}\eta_3^{n}\chi_{3'}^{n}g_{3'1'}^{\mu\ast}}{\varepsilon  - \epsilon_n - \omega_{\mu} + i\delta} 
+ \sum\limits_{\varkappa m}\frac{{\gamma}_{13}^{\varkappa(+)\ast}\chi_3^{m}\eta_{3'}^
{m}{g}_{1'3'}^{\varkappa}}{\varepsilon  + \epsilon_m + \omega_{\varkappa} - i\delta} 
\Bigr].\nonumber\\
\label{Sigma_hr}
\eea

Eqs. (\ref{Sigma_r},\ref{Sigma_1r},\ref{Sigma_2r},\ref{Sigma_hr}) are illustrated diagrammatically in Fig. \ref{SEgen}.
Notice here, that in Eqs. (\ref{Sigma_r},\ref{Sigma_1r},\ref{Sigma_2r},\ref{Sigma_hr}) we still formally distinguish between the states $|\nu\rangle$, $|\mu\rangle$ and $|\varkappa\rangle$ in systems with even particle numbers as well as between the states $|n\rangle$ and $|m\rangle$ with odd particle numbers, although the particle number constraint is already partly relaxed in the "mixed" terms containing the products of $g$ and $\gamma$ vertices (last two lines in each of Eqs. (\ref{Sigma_r},\ref{Sigma_1r},\ref{Sigma_2r},\ref{Sigma_hr})). If we further relax the particle number constraint and imply the  particle-number non-conserving approximation, as it is done in the HFB and in the quasiparticle random phase approximation (QRPA) \cite{RingSchuck1980}, Eqs. (\ref{Sigma_r},\ref{Sigma_1r},\ref{Sigma_2r},\ref{Sigma_hr}) can be combined in the superposition of Eq. (\ref{Sigma+}) as follows:
\bea
\Sigma^{r(+)}_{\nu\nu'}(\varepsilon) = \sum\limits_{\nu''\mu} \Bigl[ 
\frac{{\Gamma}^{(11)\mu}_{\nu\nu''}{\Gamma}^{(11)\mu\ast}_{\nu'\nu''}}{\varepsilon  - E_{\nu''} - \omega_{\mu} + i\delta} +
\frac{{\Gamma}^{(02)\mu\ast}_{\nu\nu''}{\Gamma}^{(02)\mu}_{\nu'\nu''}}{\varepsilon + E_{\nu''} + \omega_{\mu} - i\delta} \Bigl], \nonumber\\
\label{SEqp}
\eea
where the energies $E_{\nu''}$ marked by the Greek index stand for the quasiparticle energies of the Bogoliubov kind,
and the vertex functions $\Gamma^{(11)}$ and $\Gamma^{(02)}$ are defined as follows:
\bea
\Gamma^{(11)\mu}_{\nu\nu'} = \sum\limits_{12}\Bigl[ 
U^{\dagger}_{\nu 1}(g^{\mu}_{12}\eta^{\nu'}_2 + \gamma^{\mu(+)}_{12}\chi^{\nu'\ast}_2) - \nonumber \\
- V^{\dagger}_{\nu 1}((g^{\mu }_{12})^T\chi^{\nu'\ast}_2 + (\gamma^{\mu(-)}_{12})^T\eta^{\nu'}_2)\Bigr]
\label{Gamma11_qp}
\\
\Gamma^{(02)\mu}_{\nu\nu'} = -\sum\limits_{12}\Bigl[ 
V^{T}_{\nu 1}(g^{\mu}_{12}\eta^{\nu'}_2 + \gamma^{\mu(+)}_{12}\chi^{\nu'\ast}_2) - \nonumber \\
- U^{T}_{\nu 1}((g^{\mu }_{12})^T\chi^{\nu'\ast}_2 + (\gamma^{\mu(-)}_{12})^T\eta^{\nu'}_2)\Bigr].
\label{Gamma02_qp}
\eea
 In this way, one arrives at the forward component of the self-energy $\Sigma^{r(+)}$ (\ref{SEqp}) in the quasiparticle basis. Remarkably, the vertices $\Gamma^{(11)}$ and $\Gamma^{(02)}$ are linear combinations of the vertices of the normal and pairing phonons. From the form of Eq. (\ref{SEqp}) it is clear that in the superfluid case these phonons are components of the unified vibrations which exist simultaneously in $N$-partilce and $N\pm 2$-particle systems. This is a consequence of the particle-number non-conservation, and it is a well-known feature of approaches like QRPA. Fig. \ref{SEgen} further clarifies how such vibrations enter the components of the dynamical self-energy. 
 
 The matrix elements $\eta$ and $\chi$ in the generalized vertices $\Gamma^{(11)}$ and $\Gamma^{(02)}$, which are defined in Eqs. (\ref{etachi}), contain the information about the many-body structure of the single-fermion states. The leading approximation to the self-energy of Eq. (\ref{SEqp}) would imply the mean-field, or HFB, character of the intermediate quasiparticle states $\nu''$. In this case, the matrix elements $\eta$ and $\chi$ in Eqs. (\ref{Gamma11_qp},\ref{Gamma02_qp}) reduce to:
\bea
\eta^{\nu}_{1} = \langle 0|\psi_1|\nu \rangle = U_{1\nu}, \nonumber\\
\chi^{\nu}_{1} = \langle \nu|\psi_1|0\rangle = V^{\ast}_{1\nu},
\label{etachiHFB}
\eea
so that the vertices $\Gamma^{(11)}$ and $\Gamma^{(02)}$ reduce to:
\bea
\Gamma^{(11)\mu}_{\nu\nu'} = \sum\limits_{12}\Bigl[ 
U^{\dagger}_{\nu 1}g^{\mu}_{12}U_{2\nu'} + U^{\dagger}_{\nu 1}\gamma^{\mu(+)}_{12}V_{2\nu'} \nonumber \\
- V^{\dagger}_{\nu 1}(g^{\mu }_{12})^TV_{2\nu'} - V^{\dagger}_{\nu 1}(\gamma^{\mu(-)}_{12})^TU_{2\nu'}\Bigr]  \nonumber\\
\equiv 
\Bigl[ 
U^{\dagger}g^{\mu}U + U^{\dagger}\gamma^{\mu(+)}V 
- V^{\dagger}g^{\mu T}V - V^{\dagger}\gamma^{\mu(-)T}U\Bigr]_{\nu\nu'} \nonumber\\
\label{Gamma11_HFB}
\eea
\bea
\Gamma^{(02)\mu}_{\nu\nu'} = -\sum\limits_{12}\Bigl[ 
V^{T}_{\nu 1}g^{\mu}_{12}U_{2\nu'} + V^{T}_{\nu 1}\gamma^{\mu(+)}_{12}V_{2\nu'}  \nonumber \\
- U^{T}_{\nu 1}(g^{\mu }_{12})^TV_{2\nu'} - U^{T}_{\nu 1}(\gamma^{\mu(-)}_{12})^TU_{2\nu'}\Bigr] \nonumber\\
\equiv 
-\Bigl[ 
V^{T}g^{\mu}U+ V^{T}\gamma^{\mu(+)}V 
- U^{T}g^{\mu T}V - U^{T}\gamma^{\mu(-)T}U\Bigr]_{\nu\nu'}.\nonumber\\
\label{Gamma02_HFB}
\eea

The expression for the backward component of the fermionic self-energy in the quasiparticle basis $\Sigma^{r(-)}$ can be obtained in a similar way  by combining Eqs. (\ref{Sigma_r},\ref{Sigma_1r},\ref{Sigma_2r},\ref{Sigma_hr}) in the superposition of Eq. (\ref{Sigma-}). 
As already mentioned above, analyzing the component structure of the Dyson equation in the quasiparticle basis (\ref{Dyson_qp}) and the propagators (\ref{MFG_qp},\ref{G_qp}), it is easy to see that Eqs. (\ref{Dyson_qp}) for $(+)$ and $(-)$ components of the quasiparticle propagator have the same solutions for the energies $E_n$ and the spectroscopic factors $S^{(\pm)n}_{\nu\nu'}$, so that one of the two equations (\ref{Dyson_qp}) is redundant unless the particle number conservation is restored. Thus, summarizing the discussion of this subsection, the final Gor'kov-Dyson equation for the quasiparticle propagator in our quasiparticle-vibration coupling (QVC) approach takes the form
\be
G^{(+)}_{\nu\nu'}(\varepsilon) = {\tilde G}^{(+)}_{\nu\nu'}(\varepsilon) + \sum\limits_{\mu\mu'}{\tilde G}^{(+)}_{\nu\mu}(\varepsilon)\Sigma^{r(+)}_{\mu\mu'}(\varepsilon)G^{(+)}_{\mu'\nu'}(\varepsilon)
\label{Dyson_qp_1}
\ee
with the mean-field quasiparticle propagator ${\tilde G}^{(+)}$ defined by Eq. (\ref{MFG_qp}) and the dynamical self-energy given by Eq. (\ref{SEqp})
with the vertices of Eqs. (\ref{Gamma11_qp},\ref{Gamma02_qp}). In the leading approximation with respect to the QVC, which implies the mean-field character of the intermediate fermionic states in the self-energy, the vertices simplify to the form of Eqs. (\ref{Gamma11_HFB},\ref{Gamma02_HFB}). A more advanced approach would imply an iterative self-consistent procedure, where, after solving the Gor'kov-Dyson equation, the quasiparticle propagators are inserted back into the self-energy, and the procedure is repeated until convergence. In that case, the more general expressions of Eqs. (\ref{Gamma11_qp},\ref{Gamma02_qp}) should be employed, with the obvious correspondence between the matrix elements $\eta$ and $\chi$ and the spectroscopic factors $S^{(\pm)}$. This type of calculation scheme was implemented, for instance, in Refs. \cite{Soma2011,Soma2013,Soma2014a,Soma2021}, but in a perturbative approach to the dynamical self-energy, which does not fully include collective QVC effects.

\subsection{Strength function and transition amplitudes}
\label{SFAmpl}

As it is clear from the definitions of the QVC vertices (\ref{Gamma11_qp},\ref{Gamma02_qp},\ref{vert_pp},\ref{vert_ph}), they should be calculated by solving the equations of motion for the response function (\ref{phresp}) and the fermionic pair propagator (\ref{ppGF}). The EOM's for these functions were discussed, in particular, in Refs. \cite{LitvinovaSchuck2019,LitvinovaSchuck2020}. 
However, similarly to the one-fermion EOM, in the superfluid regime the EOM's for the particle-hole response and for the pair propagator are coupled as these propagators form the components of one object, the two-quasiparticle propagator. This becomes possible with relaxing the particle number constraint on both the ground and excited states. 

The response of a many-body fermionic system to an external probe associated with the field operator $F$ is characterized by the strength function defined as
\be
S(\omega) = \sum\limits_n \Bigl[ |\langle n|F|0\rangle |^2\delta(\omega-\omega_n) - |\langle n|F^{\dagger}|0\rangle |^2\delta(\omega+\omega_n)
\Bigr],
\label{SF}
\ee
where the summation over $n$ runs through all excited states.
As we continue to consider a more general case of superfluid systems, it is convenient to express the operator $F$ in terms of the quasiparticle fields:
\bea
F = \frac{1}{2}\sum\limits_{\mu\mu'} \Bigl(F^{20}_{\mu\mu'}\alpha^{\dagger}_{\mu}\alpha^{\dagger}_{\mu'} + 
F^{02}_{\mu\mu'}\alpha_{\mu'}\alpha_{\mu} \Bigr)\nonumber\\
F^{\dagger} = \frac{1}{2}\sum\limits_{\mu\mu'} \Bigl(F^{20\ast}_{\mu\mu'}\alpha_{\mu'}\alpha_{\mu} +
F^{02\ast}_{\mu\mu'}\alpha^{\dagger}_{\mu}\alpha^{\dagger}_{\mu'}  
\Bigr).
\label{Fext}
\eea
While Eq. (\ref{SF}) is model independent, the matrix elements in it obviously depend on the model assumptions about both the ground $|0\rangle$ and excited $|n\rangle$ states.
Let us consider here the simplest case, when the excitations are determined by the action of the one-phonon operator $Q^{n\dagger}$, such as
\bea
Q^{n\dagger} = \frac{1}{2}\sum\limits_{\mu\mu'} \Bigl(X^{n}_{\mu\mu'}\alpha^{\dagger}_{\mu}\alpha^{\dagger}_{\mu'} - 
Y^{n}_{\mu\mu'}\alpha_{\mu'}\alpha_{\mu} \Bigr) \label{Qdagger}\\
Q^{n} = \frac{1}{2}\sum\limits_{\mu\mu'} \Bigl(X^{n\ast}_{\mu\mu'}\alpha_{\mu'}\alpha_{\mu} -
Y^{n\ast}_{\mu\mu'}\alpha^{\dagger}_{\mu}\alpha^{\dagger}_{\mu'}  
 \Bigr), \label{Q}
\eea
on the ground state, while its hermitian conjugate determines the vacuum condition, i. e.,
\be
|n\rangle = Q^{n\dagger}|0\rangle \ \ \ \ \ \ Q^{n}|0\rangle = 0.
\ee
Below we discuss the QVC vertex extraction for the simplest excitation operator (\ref{Qdagger}) and its conjugate (\ref{Q}), which are just the superpositions of two-quasiparticle operators and represent the QRPA. However, it will be clear below that the approach can be further generalized to more complex excitation operators.

The amplitudes $X^{n}$ and $Y^{n}$ appearing in Eqs. (\ref{Qdagger}) and (\ref{Q}) are to be determined from the equations of motion.
The overlap of two excited state wave functions reads
\bea
\langle n|n'\rangle = \langle 0|Q^{n}Q^{n'\dagger}|0\rangle = \langle 0|[Q^{n},Q^{n'\dagger}]|0\rangle = \nonumber \\
= \sum\limits_{\mu\leq\mu'}\sum\limits_{\nu\leq\nu'}\langle 0|\Bigl[(X^{n\ast}_{\mu\mu'}\alpha_{\mu'}\alpha_{\mu} -
Y^{n\ast}_{\mu\mu'}\alpha^{\dagger}_{\mu}\alpha^{\dagger}_{\mu'}), \nonumber \\ 
(X^{n'}_{\nu\nu'}\alpha^{\dagger}_{\nu}\alpha^{\dagger}_{\nu'} - 
Y^{n'}_{\nu\nu'}\alpha_{\nu'}\alpha_{\nu} )
 \Bigr]|0\rangle = \nonumber \\
= \sum\limits_{\mu\leq\mu'}\sum\limits_{\nu\leq\nu'} \langle 0|
\Bigl(X^{n\ast}_{\mu\mu'}X^{n'}_{\nu\nu'}[\alpha_{\mu'}\alpha_{\mu},\alpha^{\dagger}_{\nu}\alpha^{\dagger}_{\nu'}] + \nonumber\\
+ Y^{n\ast}_{\mu\mu'}Y^{n'}_{\nu\nu'}[\alpha^{\dagger}_{\mu}\alpha^{\dagger}_{\mu'},\alpha_{\nu'}\alpha_{\nu}]
\Bigr)
|0\rangle .
\eea
If the commutators are evaluated in the quasiboson approximation
\bea
\langle 0|[\alpha_{\mu'}\alpha_{\mu},\alpha^{\dagger}_{\nu}\alpha^{\dagger}_{\nu'}]|0\rangle \approx
\langle HFB|[\alpha_{\mu'}\alpha_{\mu},\alpha^{\dagger}_{\nu}\alpha^{\dagger}_{\nu'}]|HFB\rangle  = \nonumber\\
= \delta_{\mu\nu}\delta_{\mu'\nu'}\nonumber\\
\langle 0|[\alpha^{\dagger}_{\mu}\alpha^{\dagger}_{\mu'},\alpha_{\nu'}\alpha_{\nu}]|0\rangle \approx
\langle HFB|[\alpha^{\dagger}_{\mu}\alpha^{\dagger}_{\mu'},\alpha_{\nu'}\alpha_{\nu}]|HFB\rangle  = \nonumber\\
= -\delta_{\mu\nu}\delta_{\mu'\nu'}, \nonumber\\
\label{QBA}
\eea
the following orthonormality relation can be obtained for the $X^{n}$ and $Y^{n}$ amplitudes:
\be
\frac{1}{2}\sum\limits_{\mu\mu'}(X^{n\ast}_{\mu\mu'}X^{n'}_{\mu\mu'} - Y^{n\ast}_{\mu\mu'}Y^{n'}_{\mu\mu'}) = \delta_{nn'}. 
\ee
 The associated completeness relations read:
\bea
\sum\limits_{n}(X^{n}_{\mu\mu'}X^{n\ast}_{\nu\nu'} - Y^{n\ast}_{\mu\mu'}Y^{n}_{\nu\nu'}) = \delta_{\mu\nu}\delta_{\mu'\nu'},\ \ \ \ \ 
\mu\leq\mu', \nu\leq\nu' \nonumber\\
\sum\limits_{n}(X^{n}_{\mu\mu'}Y^{n\ast}_{\nu\nu'} - Y^{n\ast}_{\mu\mu'}X^{n}_{\nu\nu'}) = 0, \ \ \ \ \ 
\mu\leq\mu', \nu\leq\nu'. \nonumber\\
\eea

For our purposes it is convenient to express the external field operator (\ref{Fext}) and, subsequently, the strength function (\ref{SF}), in terms of the $X^{n}$ and $Y^{n}$ amplitudes of the excitation (phonon) operators $Q^{n\dagger}$ and $Q^{n}$. With the established properties of the $X^{n}$ and $Y^{n}$ amplitudes, the pairs of the quasiparticle field operators can be then isolated by constructing the linear combinations:
\bea
\sum\limits_n(X^{n\ast}_{\mu\mu'}Q^{n\dagger} + Y^{n}_{\mu\mu'}Q^{n}) =  \alpha^{\dagger}_{\mu}\alpha^{\dagger}_{\mu'} \\
\sum\limits_n(Y^{n\ast}_{\mu\mu'}Q^{n\dagger} + X^{n}_{\mu\mu'}Q^{n}) = \alpha_{\mu'}\alpha_{\mu}.
\eea
With these relationships, the matrix elements of the external field operator from Eq. (\ref{SF}) read:
\bea
\langle n|F|0\rangle = \langle n|\sum\limits_{\mu\leq\mu'} \Bigl(F^{20}_{\mu\mu'}\alpha^{\dagger}_{\mu}\alpha^{\dagger}_{\mu'} + 
F^{02}_{\mu\mu'}\alpha_{\mu'}\alpha_{\mu} \Bigr) |0\rangle = \nonumber\\
= \sum\limits_{\mu\leq\mu'} \langle n|F^{20}_{\mu\mu'}\sum\limits_{n'}(X^{n'\ast}_{\mu\mu'}Q^{n'\dagger} + Y^{n'}_{\mu\mu'}Q^{n'}) + \nonumber\\
+F^{02}_{\mu\mu'}\sum\limits_{n'}(Y^{n'\ast}_{\mu\mu'}Q^{n'\dagger} + X^{n'}_{\mu\mu'}Q^{n'})
|0\rangle = \nonumber\\
= \sum\limits_{\mu\leq\mu'} (F^{20}_{\mu\mu'}X^{n\ast}_{\mu\mu'} + F^{02}_{\mu\mu'}Y^{n\ast}_{\mu\mu'}),\nonumber\\
\eea
\bea
\langle n|F^{\dagger}|0\rangle = \langle n|\sum\limits_{\mu\leq\mu'} \Bigl(F^{20\ast}_{\mu\mu'}\alpha_{\mu'}\alpha_{\mu} +
F^{02\ast}_{\mu\mu'}\alpha^{\dagger}_{\mu}\alpha^{\dagger}_{\mu'}  
\Bigr) |0\rangle = \nonumber\\
= \sum\limits_{\mu\leq\mu'} \langle n|F^{20\ast}_{\mu\mu'}\sum\limits_{n'}(Y^{n'\ast}_{\mu\mu'}Q^{n'\dagger} + X^{n'}_{\mu\mu'}Q^{n'}) + \nonumber\\
+ F^{02\ast}_{\mu\mu'}\sum\limits_{n'}(X^{n'\ast}_{\mu\mu'}Q^{n'\dagger} + Y^{n'}_{\mu\mu'}Q^{n'})
|0\rangle = \nonumber\\
= \sum\limits_{\mu\leq\mu'} (F^{02\ast}_{\mu\mu'}X^{n\ast}_{\mu\mu'} + F^{20\ast}_{\mu\mu'}Y^{n\ast}_{\mu\mu'}).\nonumber\\
\eea

In the practical implementations of the strength function calculation, the delta-functions in Eq. (\ref{SF}) are approximated by the Lorentz distribution
\be
\delta(\omega-\omega_n) = \frac{1}{\pi}\lim\limits_{\Delta \to 0}\frac{\Delta}{(\omega - \omega_n)^2 + \Delta^2},
\ee
so that
\bea
S(\omega) = \frac{1}{\pi}\lim\limits_{\Delta \to 0}\sum\limits_n \Bigl[ |\langle n|F|0\rangle |^2\frac{\Delta}{(\omega - \omega_n)^2 + \Delta^2}
- \nonumber \\
- |\langle n|F^{\dagger}|0\rangle |^2\frac{\Delta}{(\omega + \omega_n)^2 + \Delta^2}
\Bigr] = \nonumber\\
= -\frac{1}{\pi}\lim\limits_{\Delta \to 0} \text{Im}\sum\limits_n \Bigl[\frac{ |\langle n|F|0\rangle |^2}{\omega - \omega_n + i\Delta}
- \frac{|\langle n|F^{\dagger}|0\rangle |^2}{\omega + \omega_n + i\Delta}
\Bigr]= \nonumber\\
= -\frac{1}{\pi}\lim\limits_{\Delta \to 0} \text{Im} \Pi(\omega),\nonumber\\
\label{SFDelta} 
\eea
where $\Pi(\omega)$ is the polarizability of the system:
\bea
\Pi(\omega) = \sum\limits_n \Bigl[ \frac{|\langle n|F|0\rangle |^2}{\omega - \omega_n + i\Delta}
- \frac{|\langle n|F^{\dagger}|0\rangle |^2}{\omega + \omega_n + i\Delta}
\Bigr] = \nonumber\\
=  \sum\limits_n \Bigl[ \frac{B_n}{\omega - \omega_n + i\Delta}
- \frac{{\bar B}_n}{\omega + \omega_n + i\Delta}
\Bigr]
\label{Polar}
\eea
with the transition probabilities defined as:
\be
B_n = |\langle n|F|0\rangle |^2\ \ \ \ \ \ \ \ 
{\bar B}_n = |\langle n|F^{\dagger}|0\rangle |^2.
\label{Prob}
\ee

Notice that Eqs. (\ref{SFDelta},\ref{Polar},\ref{Prob}) are model-independent, i.e., valid for any  excitation operator and any type of ground state. For the external field operator of one-body type (\ref{Fext}), the polarizability takes the form:
\bea
\Pi(\omega) 
=  \sum\limits_n \Bigl[ \frac{\langle n|F|0\rangle
\sum\limits_{\mu\leq\mu'} (F^{20\ast}_{\mu\mu'}X^{n}_{\mu\mu'} + F^{02\ast}_{\mu\mu'}Y^{n}_{\mu\mu'})}{\omega - \omega_n + i\Delta}
- \nonumber\\
- \frac{\sum\limits_{\mu\leq\mu'} (F^{02\ast}_{\mu\mu'}X^{n\ast}_{\mu\mu'} + F^{20\ast}_{\mu\mu'}Y^{n\ast}_{\mu\mu'})\langle n|F^{\dagger}|0\rangle^{\ast} }{\omega + \omega_n + i\Delta}
\Bigr].\nonumber\\
\label{Polar1}
\eea
Grouping the parts associated with the same external field matrix elements, one realizes that the following combinations
\bea
X_{\mu\mu'}(\omega) \equiv \delta{\cal R}^{20}_{\mu\mu'}(\omega) = 
\sum\limits_n \Bigl[ \frac{X^n_{\mu\mu'}\langle n|F|0\rangle }{\omega - \omega_n + i\Delta}
- \frac{Y^{n\ast}_{\mu\mu'}\langle 0|F|n\rangle }{\omega + \omega_n + i\Delta}
\Bigr] 
\nonumber\\
Y_{\mu\mu'}(\omega) \equiv \delta{\cal R}^{02}_{\mu\mu'}(\omega) = 
\sum\limits_n \Bigl[ \frac{Y^n_{\mu\mu'}\langle n|F|0\rangle }{\omega - \omega_n + i\Delta}
- \frac{X^{n\ast}_{\mu\mu'}\langle 0|F|n\rangle }{\omega + \omega_n + i\Delta}
\Bigr] \nonumber\\
\label{XYampl}
\eea
can be used to compute the polarizability as follows:
\be
\Pi(\omega) = \sum\limits_{\mu\leq\mu'} \Bigl(F^{20\ast}_{\mu\mu'} X_{\mu\mu'}(\omega) + F^{02\ast}_{\mu\mu'} Y_{\mu\mu'}(\omega)\Bigr),
\label{Polar2}
\ee
that is consistent with Ref. \cite{Hinohara2013}. 

Eqs. (\ref{XYampl}) express the important relationship between the amplitudes $X_{\mu\mu'}(\omega), Y_{\mu\mu'}(\omega)$ and 
 $X^n_{\mu\mu'}, Y^n_{\mu\mu'}$. In the vicinity of a particular frequency, for instance, at $\omega \to \omega_n$ one term dominates the sums in Eq. (\ref{XYampl})
\bea
X_{\mu\mu'}(\omega\to \omega_n) = 
\frac{X^n_{\mu\mu'}\langle n|F|0\rangle }{\omega - \omega_n + i\Delta},
\nonumber\\
Y_{\mu\mu'}(\omega\to \omega_n) =
 \frac{Y^n_{\mu\mu'}\langle n|F|0\rangle }{\omega - \omega_n + i\Delta},
\label{XYamplvic}
\eea
so that Eqs. (\ref{XYamplvic}) can be inverted for determining the amplitudes $X^n_{\mu\mu'}, Y^n_{\mu\mu'}$ (see also Ref. \cite{Hinohara2013}):
\bea
X^n_{\mu\mu'} = \frac{1}{\langle n|F|0\rangle} \oint\limits_{\gamma_n}X_{\mu\mu'}(\omega)\frac{d\omega}{2\pi i}\nonumber\\
Y^n_{\mu\mu'} = \frac{1}{\langle n|F|0\rangle} \oint\limits_{\gamma_n}Y_{\mu\mu'}(\omega)\frac{d\omega}{2\pi i},
\label{XYampl1}
\eea
where $\gamma_n$ is a contour enclosing the pole $\omega = \omega_n - i\Delta$. Up to a phase, the matrix element $\langle n|F|0\rangle$ is determined by the transition probability $B_n$, i.e.,
\be
\langle n|F|0\rangle = e^{i\phi}\sqrt{B_n},
\ee
and the opposite phase is contained in the amplitudes $X^n_{\mu\mu'}, Y^n_{\mu\mu'}$ as they are the matrix elements of the transition density matrix $\langle 0|{\cal R}_{\mu\mu'}|n \rangle$. So, if the residues in Eq. (\ref{XYamplvic}) are real, the amplitudes $X^n_{\mu\mu'}, Y^n_{\mu\mu'}$ can be alternatively found as 
\bea
X^n_{\mu\mu'} = -\lim\limits_{\Delta \to 0}\frac{\Delta}{\sqrt{B_n}} \text{Im}X_{\mu\mu'}(\omega_n)\nonumber\\
Y^n_{\mu\mu'} = -\lim\limits_{\Delta \to 0}\frac{\Delta}{\sqrt{B_n}} \text{Im}Y_{\mu\mu'}(\omega_n),
\label{XYampl2}
\eea
up to an unimportant overall phase factor. Since the transition probabilities and the strength function values at the poles are related via Eq. (\ref{SFDelta}),
\be
B_n = \pi\lim\limits_{\Delta \to 0} \Delta\cdot S(\omega_n),
\ee
with the given strength function the $X^n$ and $Y^n$ can be alternatively found via:
\bea
X^n_{\mu\mu'} = -\lim\limits_{\Delta \to 0}\sqrt{\frac{\Delta}{\pi S(\omega_n)}} \text{Im}X_{\mu\mu'}(\omega_n)\nonumber\\
Y^n_{\mu\mu'} = -\lim\limits_{\Delta \to 0}\sqrt{\frac{\Delta}{\pi S(\omega_n)}} \text{Im}Y_{\mu\mu'}(\omega_n).
\label{XYampl3}
\eea
From the analysis of Eqs. (\ref{Polar1} -- \ref{Polar2},\ref{respspec},\ref{resppp}), see also Ref. \cite{Hinohara2013}, it is clear that the amplitudes $X^n$ and $Y^n$ are the transition densities for the transitions between the ground state and excited states $|n\rangle$ in the quasiparticle basis, while their energy-dependent counterparts $X(\omega)$ and $Y(\omega)$ are variations of the densities in the external field $F$, that is reflected in the notations used in Eq. (\ref{XYampl}). While the latter amplitudes contain the information about the external field, the former ones do not, being the solutions of the homogenius EOM's. The latter aspect will be discussed in detail in the next section. Here we notice that the coefficients between these pairs of amplitudes in Eqs. (\ref{XYampl1},\ref{XYampl2},\ref{XYampl3}) carry the  information about the external field contained in the amplitudes $X(\omega)$ and $Y(\omega)$. Notice also, that the established relationships between the pairs of amplitudes $X(\omega), Y(\omega)$ and 
$X^n, Y^n$ should be valid for the transition amplitudes beyond QRPA, when the excitation operator has a more complex structure.

\section{Phonon vertex extraction in the finite amplitude formalism}
\label{PVCvert}

Both pairs of amplitudes $X(\omega), Y(\omega)$ and 
$X^n, Y^n$ are the solutions of the QRPA equations. As mentioned above, while 
 the amplitudes $X^n, Y^n$ satisfy the QRPA equation, which does not contain explicitly the external field:
\bea
\left(\begin{array}{cc} A & B \\ B^{\ast} & A^{\ast} \end{array}\right)  \left(\begin{array}{c} X^n\\ Y^n\end{array}\right) 
 = \omega_n \left(\begin{array}{c} X^n\\ -Y^n\end{array}\right), \nonumber\\
\label{QRPA}
\eea 
the amplitudes $X(\omega), Y(\omega)$ satisfy the QRPA equation in the presence of external field 
or,
\bea
\left(\begin{array}{cc} A & B \\ B^{\ast} & A^{\ast} \end{array}\right)  \left(\begin{array}{c} X(\omega)\\ Y(\omega)\end{array}\right) +  
 \left(\begin{array}{c} F^{20}\\ F^{02}\end{array}\right) = \omega \left(\begin{array}{c} X(\omega)\\ -Y(\omega)\end{array}\right), \nonumber\\
\label{QRPAF}
\eea
sometimes called the linear response equation \cite{RingSchuck1980}. The matrices $A$ and $B$ are associated with the variations of the components of the quasiparticle Hamiltonian in the quasiparticle basis \cite{Hinohara2013},
\bea
\delta H^{20}_{\mu\mu'}(\omega) = \sum\limits_{\nu\leq\nu'} \bigl(A_{\mu\mu',\nu\nu'}X_{\nu\nu'}(\omega) + B_{\mu\mu',\nu\nu'}Y_{\nu\nu'}(\omega)\bigr) - \nonumber\\
- (E_{\mu} + E_{\mu'}) X_{\mu\mu'}(\omega) \nonumber\\
\delta H^{02}_{\mu\mu'}(\omega) = \sum\limits_{\nu\leq\nu'} \bigl(A^{\ast}_{\mu\mu',\nu\nu'}Y_{\nu\nu'}(\omega) + B^{\ast}_{\mu\mu',\nu\nu'}X_{\nu\nu'}(\omega)\bigr) -\nonumber\\
- (E_{\mu} + E_{\mu'}) Y_{\mu\mu'}(\omega),\nonumber\\
\label{dHAB}
\eea
such that
\be
\delta {H}(t) = \eta\bigl(\delta {H}(\omega)e^{-i\omega t} + \delta {H}^{\dagger}(\omega)e^{i\omega t}\bigr)
\ee
is the variation of the Hamiltonian $H(t) = H_0 + \delta H(t)$ in response to small oscillations of an external field
\be
{F}(t) = \eta({F}e^{-i\omega t} + { F}^{\dagger}e^{i\omega t}),
\ee
with $\eta$ being a small linear expansion parameter and $H_0$ the mean-field Hamiltonian of the Bogoliubov quasiparticles:
\bea
H_0 = \left( \begin{array}{cc}h - \lambda & \Delta \\ -\Delta^{\ast}  & -h^{\ast} + \lambda \end{array} \right),
\eea
where $\lambda$ is the chemical potential.
Eq. (\ref{dHAB}) leads to the finite-amplitude form of the QRPA equation (\ref{QRPAF}):
\bea
X_{\mu\nu}(\omega)  = \frac{\delta{\cal H}^{20}_{\mu\nu}(\omega) + F^{20}_{\mu\nu}}{\omega - E_{\mu} - E_{\nu}} \nonumber\\
Y_{\mu\nu}(\omega)  = \frac{\delta{\cal H}^{02}_{\mu\nu}(\omega) + F^{02}_{\mu\nu}}{-\omega - E_{\mu} - E_{\nu}}.\nonumber\\
\label{FAM}
\eea
The full variation of the quasiparticle Hamiltonian $\delta H(\omega)$ has the following matrix structure:
\bea
\delta {H}(\omega) = \frac{1}{2}\Bigl(\begin{array}{cc} \alpha^{\dagger} & \alpha \end{array}\Bigr)
\Bigl(\begin{array}{cc} \delta H^{11}(\omega) & \delta H^{20}(\omega) \\ -\delta H^{02}(\omega) & -\delta H^{11T}(\omega)\end{array}\Bigr)
\Bigl(\begin{array}{c} \alpha \\ \alpha^{\dagger} \end{array}\Bigr),\nonumber\\
\eea
where
\bea
\Bigl(\begin{array}{cc} \delta H^{11}(\omega) & \delta H^{20}(\omega) \\ -\delta H^{02}(\omega) & -\delta H^{11T}(\omega)\end{array}\Bigr) = 
\nonumber\\
= {\cal W}^{\dagger}\Bigl(\begin{array}{cc} \delta h(\omega) & \delta\Delta^{(+)}(\omega) \\ -\delta\Delta^{(-)\ast}(\omega) & -\delta h^T(\omega)\end{array}\Bigr){\cal W} =\nonumber\\
= \left( \begin{array}{cc} U^{\dagger} & V^{\dagger} \\ V^T & U^T \end{array} \right)
\Bigl(\begin{array}{cc} \delta h(\omega) & \delta\Delta^{(+)}(\omega) \\ -\delta\Delta^{(-)\ast}(\omega) & -\delta h^T(\omega)\end{array}\Bigr)
\left( \begin{array}{cc} U & V^{\ast} \\ V & U^{\ast} \end{array} \right),\nonumber\\
\eea
so that the components of the variation $\delta { H}$ are expressed through the variations of the mean-field single-particle Hamiltonian $h$ and pairing fields  $\Delta^{(\pm)}$:
\bea
\delta H^{11} &=& U^{\dagger}\delta h U + U^{\dagger}\delta\Delta^{(+)}V - V^{\dagger}\delta\Delta^{(-)\ast}U - \nonumber\\
&-& V^{\dagger}\delta h^T V\nonumber\\
\delta H^{20} &=& U^{\dagger}\delta h V^{\ast} + U^{\dagger}\delta\Delta^{(+)}U^{\ast} - V^{\dagger}\delta\Delta^{(-)\ast}V^{\ast} - \nonumber\\
&-&V^{\dagger}\delta h^T U^{\ast}\nonumber\\
-\delta H^{02} &=& V^{T}\delta h U + V^{T}\delta\Delta^{(+)}V - U^{T}\delta\Delta^{(-)\ast}U - \nonumber\\
&-& U^{T}\delta h^T V\nonumber\\
-\delta H^{11T} &=& V^{T}\delta h V^{\ast} + V^{T}\delta\Delta^{(+)}U^{\ast} - U^{T}\delta\Delta^{(-)\ast}V^{\ast} - \nonumber\\
&-& U^{T}\delta h^T U^{\ast}.\nonumber\\
\label{dH}
\eea
The variations of the single-particle Hamiltonian and the pairing fields are related to the variations of the normal and pairing densities \footnote{In this section it is implied that $\bar v$ can be approximated by some effective interaction}:
\bea
\delta h_{12}(\omega) = \sum\limits_{34} {\bar v}_{1423}\delta\rho_{34}(\omega)\\
\delta\Delta^{(\pm)}_{12}(\omega) = \frac{1}{2} \sum\limits_{34} {\bar v}_{1234}\delta\varkappa^{(\pm)}_{34}(\omega),
\eea
while
the latter density variations are, in turn, related to the $X_{\mu\mu'}(\omega)$ and $Y_{\mu\mu'}(\omega)$ amplitudes \cite{Kortelainen2015,Bjelcic2020}:
\bea
\delta\rho_{12}(\omega) =  (UX(\omega)V^T + V^{\ast}Y^T(\omega)U^{\dagger})_{12}\\
\delta\rho^{\dagger}_{12}(\omega) =  (V^{\ast}X^{\dagger}(\omega)U^{\dagger} + UY^{\ast}(\omega)V^T)_{12}\\
\delta\varkappa^{(+)}_{12}(\omega) = (UX(\omega)U^T + V^{\ast}Y^T(\omega)V^{\dagger})_{12}\\
\delta\varkappa^{(-)}_{12}(\omega) = (V^{\ast}X^{\dagger}(\omega)V^{\dagger} + UY^{\ast}(\omega)U^T)_{12}.
\eea
The latter means that in the vicinity of a pole $\omega \to \omega_n$ these density variations can be related to the respective transition densities 
in the same way as $X(\omega), Y(\omega)$ are related to $X^n, Y^n$ in Eq. (\ref{XYamplvic}):
\bea
\delta\rho_{12}(\omega\to \omega_n) =
\frac{\rho^n_{12}\langle n|F|0\rangle }{\omega - \omega_n + i\Delta}\nonumber\\
\delta\varkappa^{(+)}_{12}(\omega\to \omega_n) =  
\frac{\varkappa^{n(+)}_{12}\langle n|F|0\rangle }{\omega - \omega_n + i\Delta}\nonumber\\ 
\delta\varkappa^{(-)\ast}_{12}(\omega\to \omega_n) =  
\frac{\varkappa^{n(-)\ast}_{12}\langle n|F|0\rangle }{\omega - \omega_n + i\Delta},\nonumber\\ 
\eea
where
\bea
\rho^n_{12} =  (UX^nV^T + V^{\ast}Y^{nT}(\omega)U^{\dagger})_{12}\nonumber\\
\varkappa^{n(+)}_{12} = (UX^nU^T + V^{\ast}Y^{nT}V^{\dagger})_{12}\nonumber\\
\varkappa^{n(-)}_{12} = (V^{\ast}X^{n\dagger}V^{\dagger} + UY^{n\ast}U^T)_{12}.
\eea
Now the variations in Eqs. (\ref{dH}) can be related to the latter transition densities. Namely, the variation of the single-particle Hamiltonian at $\omega \to \omega_n$ reads:
\bea
\delta h_{12}(\omega\to\omega_n) = \sum\limits_{34}\frac{\delta h_{12}(\omega)}{\delta\rho_{34}(\omega)}\delta\rho_{34}(\omega) = \nonumber\\ = \sum\limits_{34}\frac{\delta h_{12}(\omega)}{\delta\rho_{34}(\omega)}\rho^n_{34}\frac{\langle n|F|0\rangle }{\omega - \omega_n + i\Delta}=
\nonumber\\
= \sum\limits_{34}{\bar v}_{1423}\rho^n_{34}\frac{\langle n|F|0\rangle }{\omega - \omega_n + i\Delta} = \nonumber \\
= g^n_{12}\frac{\langle n|F|0\rangle }{\omega - \omega_n + i\Delta},
\label{dh}
\eea
where the definition of the phonon vertex $g^n$ (\ref{vert_ph}) was applied. Analogously, for $\delta\Delta^{(+)}$,
\bea
\delta \Delta^{(+)}_{12}(\omega\to\omega_n) = \sum\limits_{34}\frac{\delta\Delta^{(+)}_{12}(\omega)}{\delta\varkappa^{(+)}_{34}(\omega)}\delta\varkappa^{(+)}_{34}(\omega) = \nonumber\\ = \sum\limits_{34}\frac{\delta\Delta^{(+)}_{12}(\omega)}{\delta\varkappa^{(+)}_{34}(\omega)}\varkappa^{n(+)}_{34}\frac{\langle n|F|0\rangle }{\omega - \omega_n + i\Delta}=
\nonumber\\
= \frac{1}{2}\sum\limits_{34}{\bar v}_{1234}\varkappa^{n(+)}_{34}\frac{\langle n|F|0\rangle }{\omega - \omega_n + i\Delta} = \nonumber \\
= \gamma^{n(+)}_{12}\frac{\langle n|F|0\rangle }{\omega - \omega_n + i\Delta},
\label{dkp}
\eea
and, for $\delta\Delta^{(-)\ast}$,
\bea
\delta \Delta^{(-)\ast}_{12}(\omega\to\omega_n) = \sum\limits_{34}\frac{\delta\Delta^{(-)\ast}_{12}(\omega)}{\delta\varkappa^{(-)\ast}_{34}(\omega)}\delta\varkappa^{(-)\ast}_{34}(\omega) = \nonumber\\ = \sum\limits_{34}\frac{\delta\Delta^{(-)\ast}_{12}(\omega)}{\delta\varkappa^{(-)\ast}_{34}(\omega)}\varkappa^{n(-)\ast}_{34}\frac{\langle n|F|0\rangle }{\omega - \omega_n + i\Delta}=
\nonumber\\
= \frac{1}{2}\sum\limits_{34}{\bar v}^{\ast}_{1234}\varkappa^{n(-)\ast}_{34}\frac{\langle n|F|0\rangle }{\omega - \omega_n + i\Delta} = \nonumber \\
= \gamma^{n(-)T}_{12}\frac{\langle n|F|0\rangle }{\omega - \omega_n + i\Delta},\nonumber\\
\label{dkm}
\eea
where we applied the definitions of Eq. (\ref{vert_pp}) for the pairing phonon vertices $\gamma^{n(\pm)}$ and identified the pairing transition densities $\varkappa^{n(\pm)}$ with the matrix elements in the residues of the particle-particle propagator (\ref{resppp},\ref{alphabeta}) as follows:
\bea
\alpha^n_{12} = \langle 0|\psi_2\psi_1|n\rangle = \varkappa^{n(+)}_{12} \\
\beta^n_{12} = \langle 0|\psi^{\dagger}_2\psi^{\dagger}_1|n\rangle = \varkappa^{n(-)\ast}_{21}.
\label{pairden}
\eea
Now, inserting Eqs. (\ref{dh},\ref{dkp},\ref{dkm}) into Eqs. (\ref{dH}), we get at the pole $\omega \to \omega_n$
\bea
\delta H^{ij}_{\mu\mu'}(\omega \to \omega_n) = \Gamma^{(ij)n}_{\mu\mu'} \frac{\langle n|F|0\rangle }{\omega - \omega_n + i\Delta},
\label{dHGamma}
\eea
with $\{ij\} = \{11,02,20\}$, i.e., the relationship between the variations of the  quasiparticle Hamiltonian and the quasiparticle-phonon vertices, two of which were defined by Eqs. (\ref{Gamma11_HFB},\ref{Gamma02_HFB}). Notice the analogy of Eq. (\ref{dHGamma}) to Eqs. (\ref{XYamplvic}). The QVC vertices $\Gamma^{(11)n}$ and $\Gamma^{(02)n}$, which enter the quasiparticle self-energy of Eq. (\ref{SEqp}), can thus be extracted from the variations of the quasiparticle Hamiltonian $\delta H^{11}$ and $\delta H^{02}$as follows: 
\be
\Gamma^{(ij)n}_{\mu\mu'}  = \frac{1}{\langle n|F|0\rangle} \oint\limits_{\gamma_n}\delta H^{ij}_{\mu\mu'}(\omega)\frac{d\omega}{2\pi i}
\label{Gijcont}
\ee
where $\gamma_n$ is a contour enclosing the pole $\omega = \omega_n - i\Delta$. For the real residues in Eq. (\ref{dHGamma}), up to an unimportant phase, this can be simplified to 
\bea
\Gamma^{(ij)n}_{\mu\mu'} = -\lim\limits_{\Delta \to 0}\frac{\Delta}{\sqrt{B_n}} \text{Im}\delta H^{ij}_{\mu\mu'}(\omega_n + i\Delta) = \nonumber\\
= -\lim\limits_{\Delta \to 0}\sqrt{\frac{\Delta}{\pi S(\omega_n)}} \text{Im}\delta H^{ij}_{\mu\mu'}(\omega_n + i\Delta).
\label{Gijim}
\eea
The relationships (\ref{Gijcont},\ref{Gijim}) can be useful if the finite amplitude method of solving QRPA equations is available. This idea was realized in Ref. \cite{Zhang2021}, where the Gor'kov-Dyson equation for quasiparticle propagators was solved numerically on the base of the relativistic HFB. In particular, in Ref. \cite{Zhang2021} the method of the QVC vertex extraction was proven efficient enough to tackle heavy nuclei with axial deformation. Notice here that Eqs. (\ref{Gijcont},\ref{Gijim}) allow for generalization of the FAM-QRPA to the inclusion of the QVC effects in a fully variational form. This possibility will be considered elsewhere.


\section{Summary}
\label{summary}

We presented a theoretical framework for fermionic propagators in strongly-coupled superfluid fermionic many-body systems. Starting from the general Hamiltonian with a bare two-fermion interaction, we worked out the equations of motion for the Gor'kov set of two normal and two anomalous propagators. In contrast to the original Gor'kov theory, the kernels of the obtained EOM's contain dynamical components with three-fermion propagators. These propagators are approximated by factorizing them into the all possible products of two-fermion and one-fermion propagators with the relaxed particle number conservation condition. The resulting set of coupled equations called Gor'kov-Dyson equations is formulated first in the mean-field single-particle basis, i.e., the basis which diagonalizes the one-body part of the underlying Hamiltonian.   

Then, it is shown explicitly that, by the transformation to the HFB basis, the four pairwise coupled Gor'kov-Dyson equations reduce to one equation for the forward or for the dual backward component of the quasiparticle propagator. This scales down considerably the computational effort and reveals important relationships between the two representations. The dynamical kernel of the resulting equation is mapped to the quasiparticle-vibration coupling, where the normal and pairing phonons become components of the unified phonons, which can be found by solving a EOM for the two-quasiparticle fermionic propagator. Although we did not discuss the latter EOM in detail, we considered the simplest case of it known as QRPA and pointed out how the solutions of the QRPA equation can be used to extract the quasiparticle-vibration coupling vertices, which enter the dynamical kernel of the Gor'kov-Dyson equation in the quasipartice basis. Finally, we proposed a method for extracting the vertices from the QRPA implemented within the finite amplitude method. This may be particularly useful for practitioners dealing with systems with large number of fermions, such as medium-mass and heavy atomic nuclei, and non-spherical shapes. The latter case is especially difficult for traditional QRPA solvers, and FAM-QRPA and its possible extensions are the methods of choice for such systems. A recent example can be found in Ref. \cite{Zhang2021}. Therefore, the link between FAM-QRPA and the Gor'kov-Dyson equation in the quasipartice basis found in this work makes it possible to perform efficient calculations beyond the HFB mean-field approach.
It also paves the way to the QVC extensions of the FAM-QRPA. 

\section*{Acknowledgement}
This work is supported by the US-NSF Career Grant PHY-1654379.
%

%
\section*{Appendix: The Gor'kov theory as one-fermion expansion of the first EOM}
\label{appA}

The Gor'kov theory for the fermionic propagator can be obtained starting from the first EOM (\ref{spEOM}):
\be
(i\partial_t - \varepsilon_1)G_{11'}(t-t') = \delta_{11'}\delta(t-t') + R_{11'}(t-t'),
\label{spEOMrep}
\ee
where
\bea
R_{11'}(t-t') &=& i\langle T[V,\psi_1](t){\psi^{\dagger}}_{1'}(t')\rangle = \nonumber\\
&=&\frac{i}{2}\sum\limits_{ikl}{\bar v}_{i1kl}
\langle T(\psi^{\dagger}_i\psi_l\psi_k)(t){\psi^{\dagger}}_{1'}(t')\rangle.\nonumber\\
\label{Rkern}
\eea
Instead of generating the second EOM as it is done in Section \ref{normal}, Eq. (\ref{EOMR}), one can approximate the expectation value in $R_{11'}(t-t')$ by the cluster decomposition, or factorization, as
\bea
i\langle T(\psi^{\dagger}_i\psi_l\psi_k)(t){\psi^{\dagger}}_{1'}(t')\rangle \approx i\langle T\psi^{\dagger}_i(t){\psi^{\dagger}}_{1'}(t')\rangle
\langle\psi_l\psi_k\rangle - \nonumber \\
- i\langle T\psi_l(t){\psi^{\dagger}}_{1'}(t')\rangle\langle\psi^{\dagger}_i\psi_k\rangle  + 
i\langle T\psi_k(t){\psi^{\dagger}}_{1'}(t')\rangle\langle\psi^{\dagger}_i\psi_l\rangle = \nonumber \\
= -\varkappa_{kl}F^{(2)}_{i1'}(t-t')+ \rho_{ki}G_{l1'}(t-t') - \rho_{li}G_{k1'}(t-t'),\nonumber \\
\label{decom}
\eea 
with the static normal $\rho_{kl} = \langle\psi^{\dagger}_l\psi_k\rangle$ and pairing $\varkappa_{kl} = \langle\psi_l\psi_k\rangle$ densities and the anomalous fermionic propagator 
\bea
F^{(2)}_{11'}(t-t') = -i\langle T\psi^{\dagger}_1(t){\psi^{\dagger}}_{1'}(t')\rangle,
\label{F2a}
\eea
which has non-vanishing contribution if the particle number conservation is relaxed in the ground state wave function, as well as the pairing densities. As a reminder, the densities listed above are related to the components of the single-fermion self-energy  in the superfluid mean-field approximation as
\bea
{\tilde\Sigma}_{12} = \sum\limits_{kl}{\bar v}_{1k2l}\rho_{lk}  \ \ \ \ \ \ \ 
{\Delta}_{12} = \frac{1}{2}\sum\limits_{kl}{\bar v}_{12kl}\varkappa_{kl} . 
\eea
Thus, in the approximation of Eq. (\ref{decom}) the EOM for the fermionic propagator takes the form:
\bea
(i\partial_t - \varepsilon_1)G_{11'}(t-t') &=& \delta_{11'}\delta(t-t') + 
\label{spEOMrep1}
\\
&+& \sum\limits_{i}\bigl[\Delta_{1i}F^{(2)}_{i1'}(t-t') + {\tilde\Sigma}_{1i}G_{i1'}(t-t')\bigr].\nonumber
\eea
Introducing the Fourier images for the time-dependent entities of Eq. (\ref{spEOMrep1}) as
\bea
G_{11'}(t-t') &=& \int\limits_{-\infty}^{\infty} \frac{d\omega}{2\pi} e^{-i\omega(t-t')}G_{11'}(\omega)\\
F^{(2)}_{11'}(t-t') &=& \int\limits_{-\infty}^{\infty} \frac{d\omega}{2\pi} e^{-i\omega(t-t')}F^{(2)}_{11'}(\omega)\\
\delta(t-t') &=& \int\limits_{-\infty}^{\infty} \frac{d\omega}{2\pi} e^{-i\omega(t-t')},
\eea
the EOM for the fermionic propagator can be transferred to the energy (frequency) domain:
\be
G_{11'}(\omega) = G^0_{11'}(\omega) + \sum\limits_{ij}G^0_{1i}(\omega)\bigl[\Delta_{ij}F^{(2)}_{j1'}(\omega) + {\tilde\Sigma}_{ij}G_{j1'}(\omega)\bigr].
\label{EOMGe}
\ee
Thereby, one can see that the fermionic propagator $G(\omega)$, also called normal propagator in the case of superfluidity, becomes coupled to the anomalous propagator $F^{(2)}(\omega)$ via the pairing mean field $\Delta$. This indicates that, in order to have a closed system of equations, another EOM should be generated for this new propagator in the same approximation. Taking the time derivative of Eq. (\ref{F2a}) yields 
\be
(i\partial_t + \varepsilon_1)F^{(2)}_{11'}(t-t') = R^{(2)}_{11'}(t-t'),
\label{EOMF2}
\ee
with
\bea
R^{(2)}_{11'}(t-t') &=& i\langle T[V,\psi^{\dagger}_1](t){\psi^{\dagger}}_{1'}(t')\rangle = \nonumber\\
&=&\frac{i}{2}\sum\limits_{ikl}{\bar v}_{ik1l}
\langle T(\psi^{\dagger}_i\psi^{\dagger}_k\psi_l)(t){\psi^{\dagger}}_{1'}(t')\rangle.\nonumber\\
\label{R2kern}
\eea
The factorization of the expectation value in the latter term into products of one-fermion correlation functions, similarly to the case of Eq. (\ref{decom1}), leads to
\bea
i\langle T(\psi^{\dagger}_i\psi^{\dagger}_k\psi_l)(t){\psi^{\dagger}}_{1'}(t')\rangle \approx i\langle T\psi^{\dagger}_i(t){\psi^{\dagger}}_{1'}(t')\rangle
\langle\psi^{\dagger}_k\psi_l\rangle - \nonumber \\
- i\langle T\psi^{\dagger}_k(t)\psi^{\dagger}_{1'}(t')\rangle\langle\psi^{\dagger}_i\psi_l\rangle  + 
i\langle T\psi_l(t)\psi^{\dagger}_{1'}(t')\rangle\langle\psi^{\dagger}_i\psi^{\dagger}_k\rangle = \nonumber \\
= -\rho_{lk}F^{(2)}_{i1'}(t-t') + \rho_{li}F^{(2)}_{k1'}(t-t') - \varkappa^{\ast}_{ik}G_{l1'}(t-t'),\nonumber \\
\label{decom2}
\eea 
so that
\be
(i\partial_t + \varepsilon_1)F^{(2)}_{11'}(t-t') = -\sum\limits_{i}\bigl[{\tilde\Sigma}_{i1}F^{(2)}_{i1'}(t-t') + {\Delta}^{\ast}_{1i}G_{i1'}(t-t')\bigr].
\label{EOMF2a}
\ee
The Fourier transform of the latter equation reads:
\be
F^{(2)}_{11'}(\omega) = -\sum_{ij} G^{(h)0}_{1i}(\omega)\bigl[{\tilde\Sigma}^T_{ij}F^{(2)}_{j1'}(\omega) + {\Delta}^{\ast}_{ij}G_{j1'}(\omega)
\bigr].
\label{EOMF2e}
\ee

The complete set of the Gorkov's propagators includes the other two correlation functions, namely
\bea
G^{(h)}_{11}(t-t') = -i\langle T\psi^{\dagger}_1(t)\psi_{1'}(t')\rangle,\\ 
\label{Gha}
F^{(1)}_{11'}(t-t') = -i\langle T\psi_1(t)\psi_{1'}(t')\rangle .
\label{F1a}
\eea
Proceeding similarly with those, one gets:
\be
(i\partial_t + \varepsilon_1)G^{(h)}_{11'}(t-t') = \delta_{11'}\delta(t-t') + R^{(h)}_{11'}(t-t'),
\label{spEOMreph}
\ee
\bea
R^{(h)}_{11'}(t-t') &=& i\langle T[V,\psi^{\dagger}_1](t)\psi_{1'}(t')\rangle = \nonumber\\
&=&\frac{i}{2}\sum\limits_{ikl}{\bar v}_{ik1l}
\langle T(\psi^{\dagger}_i\psi^{\dagger}_k\psi_l)(t)\psi_{1'}(t')\rangle,\nonumber\\
\label{Rhkern}
\eea
\bea
i\langle T(\psi^{\dagger}_i\psi^{\dagger}_k\psi_l)(t)\psi_{1'}(t')\rangle \approx i\langle T\psi^{\dagger}_i(t)\psi_{1'}(t')\rangle
\langle\psi_k^{\dagger}\psi_l\rangle - \nonumber \\
- i\langle T\psi^{\dagger}_k(t)\psi_{1'}(t')\rangle\langle\psi^{\dagger}_i\psi_l\rangle  + 
i\langle T\psi_l(t)\psi_{1'}(t')\rangle\langle\psi^{\dagger}_i\psi^{\dagger}_k\rangle = \nonumber \\
= -\rho_{lk}G^{(h)}_{i1'}(t-t')+\rho_{li}G^{(h)}_{k1'}(t-t') - \varkappa^{\ast}_{ik}F^{(1)}_{l1'}(t-t'),\nonumber \\
\label{decomh}
\eea 
\bea
(i\partial_t &+& \varepsilon_1)G^{(h)}_{11'}(t-t') = \delta_{11'}\delta(t-t') - 
\label{spEOMreph_a}
\\
&-& \sum\limits_{i}\bigl[{\tilde\Sigma}^T_{1i}G^{(h)}_{i1'}(t-t') + \Delta^{\ast}_{1i}F^{(1)}_{i1'}(t-t')\bigr],\nonumber
\eea
\bea
G^{(h)}_{11'}(\omega) &=& G^{(h)0}_{11'}(\omega) - \nonumber\\ &-& 
\sum\limits_{ij}G^{(h)0}_{1i}(\omega)\bigl[{\tilde\Sigma}^T_{ij}G^{(h)}_{j1'}(\omega) + \Delta^{\ast}_{ij}F^{(1)}_{j1'}(\omega)\bigr]
\nonumber\\
\label{EOMGhe}
\eea
and 
\be
(i\partial_t - \varepsilon_1)F^{(1)}_{11'}(t-t') = R^{(1)}_{11'}(t-t'),
\label{EOMF1}
\ee
\bea
R^{(1)}_{11'}(t-t') &=& i\langle T[V,\psi_1](t){\psi}_{1'}(t')\rangle = \nonumber\\
&=&\frac{i}{2}\sum\limits_{ikl}{\bar v}_{i1kl}
\langle T(\psi^{\dagger}_i\psi_l\psi_k)(t)\psi_{1'}(t')\rangle,\nonumber\\
\label{R1kern}
\eea
\bea
i\langle T(\psi^{\dagger}_i\psi_l\psi_k)(t)\psi_{1'}(t')\rangle \approx i\langle T\psi^{\dagger}_i(t)\psi_{1'}(t')\rangle
\langle\psi_l\psi_k\rangle - \nonumber \\
- i\langle T\psi_l(t)\psi_{1'}(t')\rangle\langle\psi^{\dagger}_i\psi_k\rangle  + 
i\langle T\psi_k(t)\psi_{1'}(t')\rangle\langle\psi^{\dagger}_i\psi_l\rangle = \nonumber \\
= -\rho_{li}F^{(1)}_{k1'}(t-t') + \rho_{ki}F^{(1)}_{l1'}(t-t') - \varkappa_{kl}G^{(h)}_{i1'}(t-t'),\nonumber \\
\label{decom1}
\eea 
\be
(i\partial_t - \varepsilon_1)F^{(1)}_{11'}(t-t') = \sum\limits_{i}\bigl[{\tilde\Sigma}_{1i}F^{(1)}_{i1'}(t-t') + {\Delta}_{1i}G^{(h)}_{i1'}(t-t')\bigr],
\label{EOMF1a}
\ee
\be
F^{(1)}_{11'}(\omega) = \sum_{ij} G^{0}_{1i}(\omega)\bigl[{\tilde\Sigma}_{ij}F^{(1)}_{j1'}(\omega) + {\Delta}_{ij}G^{(h)}_{j1'}(\omega)
\bigr].
\label{EOMF1e}
\ee
The combination of Eqs. (\ref{EOMGe},\ref{EOMGhe},\ref{EOMF1e},\ref{EOMF2e})
\bea
G_{11'}(\omega) &=& G^0_{11'}(\omega) + \sum\limits_{ij}G^0_{1i}(\omega)\bigl[\Delta_{ij}F^{(2)}_{j1'}(\omega) + {\tilde\Sigma}_{ij}G_{j1'}(\omega)\bigr] 
\nonumber\\
F^{(2)}_{11'}(\omega) &=& -\sum_{ij} G^{(h)0}_{1i}(\omega)\bigl[{\tilde\Sigma}^T_{ij}F^{(2)}_{j1'}(\omega) + {\Delta}^{\ast}_{ij}G_{j1'}(\omega)
\bigr] \nonumber \\
G^{(h)}_{11'}(\omega) &=& G^{(h)0}_{11'}(\omega) - 
\sum\limits_{ij}G^{(h)0}_{1i}(\omega)\bigl[{\tilde\Sigma}^T_{ij}G^{(h)}_{j1'}(\omega) + \Delta^{\ast}_{ij}F^{(1)}_{j1'}(\omega)\bigr]
\nonumber\\
F^{(1)}_{11'}(\omega) &=& \sum_{ij} G^{0}_{1i}(\omega)\bigl[{\tilde\Sigma}_{ij}F^{(1)}_{j1'}(\omega) + {\Delta}_{ij}G^{(h)}_{j1'}(\omega)
\bigr]
\label{Gorkov}
\eea
constitutes the famous Gor'kov theory and describes a superfluid many-fermion system in the mean-field approximation. Eqs. (\ref{Gorkov}) can also be written in the matrix form 
\bea
\left(\begin{array}{cc}G_{11'}(\omega) & F^{(1)}_{11'}(\omega) \\ F^{(2)}_{11'}(\omega) & G^{(h)}_{11'}(\omega)\end{array}\right) = \left(\begin{array}{cc}G^0_{11'}(\omega) & 0 \\ 0 & G^{(h)0}_{11'}(\omega)\end{array}\right) + \nonumber\\
+ \sum\limits_{22'}\left(\begin{array}{cc}G^0_{12}(\omega) & 0 \\ 0 & G^{(h)0}_{12}(\omega)\end{array}\right)\left(\begin{array}{cc}{\tilde\Sigma}_{22'}(\omega) & \Delta_{22'}(\omega) \\ -\Delta^{\ast}_{22'}(\omega) & -{\tilde\Sigma}^{T}_{22'}(\omega) \end{array}\right)\times\nonumber\\
\times \left(\begin{array}{cc}G_{2'1'}(\omega) & F^{(1)}_{2'1'}(\omega) \\ F^{(2)}_{2'1'}(\omega) & G^{(h)}_{2'1'}(\omega)\end{array}\right),\nonumber\\
\label{DGEmf}
\eea
which is consistent with Eq. (\ref{DGE}) if the self-energy is confined by its mean-field part. 

\bibliography{Bibliography_Aug2021}
\end{document}